\documentclass[paper]{agujournal2019}
\usepackage{float}
\usepackage{url}
\usepackage{lineno}
\usepackage{multirow}
\usepackage{makecell}
\usepackage{changepage}
\usepackage{caption}
\usepackage{color}
\usepackage{booktabs}
\usepackage{amsmath}
\usepackage{amssymb}
\usepackage{graphicx}
\usepackage[export]{adjustbox}
\usepackage{booktabs,subcaption,amsfonts,dcolumn}
\draftfalse

\newcommand{\ssr}{   {Space Sci. Rev. }}

\newcommand{\jgr}{   {J. Geophys. Res. (Space Physics)}}
\newcommand{\grl}{   {Geophys. Res. Lett.}}

\newcommand{\apjl}{   {Astrophys. J. Lett.}}

\newcommand{\nat}{   {Nature}}
\newcommand{\aap}{   {Astronomy \& Astrophysics}}

\newcommand{\blue}{\textcolor{black}}

\journalname{JGR: Space Physics}

\begin{document}

%% ---------------------------------------------------------------%%

\title{Omnidirectional Energetic Electron Fluxes from 150 km to 20,000 km: an ELFIN-Based Model}

\authors{Emile Saint-Girons\affil{1,2}, Xiao-Jia Zhang\affil{1,3}, Didier Mourenas\affil{4,5}, Anton V. Artemyev\affil{3}, Vassilis Angelopoulos\affil{3}}
\affiliation{1}{Department of Physics, University of Texas at Dallas, Richardson, Texas, USA}
\affiliation{2}{CentraleSupélec, Gif-sur-Yvette, France}
\affiliation{3}{Department of Earth, Planetary, and Space Sciences, University of California, Los Angeles, Los Angeles, California, USA}
\affiliation{4}{CEA, DAM, DIF, Arpajon, France}
\affiliation{5}{Laboratoire Matière en Conditions Extrêmes, Université Paris-Saclay, CEA, Bruyères-le-Châtel, France}

\correspondingauthor{Xiao-Jia Zhang}{xjzhang@utdallas.edu}

\begin{keypoints}
\item Quasi-linear theory is used to infer omnidirectional electron flux along magnetic field lines from low-altitude spacecraft measurements
\item The obtained analytical model of average omnidirectional electron flux is consistent with equatorial measurements from the Van Allen Probes
\item The model shows the impact of impulsive and time-integrated \blue{substorm activity} on electron fluxes in plasma sheet and outer radiation belt
\end{keypoints}

\begin{abstract}
The strong variations of energetic electron fluxes in the Earth's inner magnetosphere are notoriously hard to forecast. Developing accurate empirical models of electron fluxes from low to high altitudes at all latitudes is therefore useful to improve our understanding of flux variations and to assess radiation hazards for spacecraft systems. In the present work, energy- and pitch-angle-resolved precipitating, trapped, and backscattered electron fluxes measured at low altitude by Electron Loss and Fields Investigation (ELFIN) CubeSats are used to infer omnidirectional fluxes at altitudes below and above the spacecraft, from 150 km to 20,000 km, making use of adiabatic transport theory and quasi-linear diffusion theory. The inferred fluxes are fitted as a function of selected parameters using a stepwise multivariate optimization procedure, providing an analytical model of omnidirectional electron flux along each geomagnetic field line, based on measurements from only one spacecraft in low Earth orbit. The modeled electron fluxes are provided as a function of $L$-shell, altitude, energy, and two different indices of past \blue{substorm activity}, computed over the preceding 4 hours or 3 days, potentially allowing to disentangle impulsive processes (such as rapid injections) from cumulative processes (such as inward radial diffusion and wave-driven energization). The model is validated through comparisons with equatorial measurements from the Van Allen Probes, demonstrating the broad applicability of the present method. The model indicates that both impulsive and time-integrated \blue{substorm activity} partly control electron fluxes in the outer radiation belt and in the plasma sheet.
\end{abstract}

\section{Introduction}

The high variability of electron fluxes trapped along geomagnetic field lines in the Earth's inner magnetosphere has been a focus of intense research since the discovery of the radiation belts, both to improve our fundamental understanding of the space environment and as a practical necessity to mitigate space weather hazards for satellites \cite{Li&Hudson19, Zheng19}. Energetic electron fluxes in the outer radiation belt (at McIlwain shells $L\geq 3-4$) vary both spatially and temporally, exhibiting steep increases after prolonged periods of high \blue{substorm activity} corresponding to plasma sheet injections accompanied by wave-driven electron energization \cite{Hua22b, Mourenas19:Impact, Mourenas23:jgr:upper_limit}, as well as sudden dropouts mainly caused by solar wind dynamic pressure impulses and magnetopause shadowing \cite{Boynton17, Shprits06:magnetopause}. Electron precipitation through resonant interactions with whistler-mode waves or electromagnetic ion cyclotron (EMIC) waves can lead to fast losses \cite{Mourenas17, Ross21:emic}, while electron energization by whistler-mode chorus waves, or through radial transport by ULF waves, can increase trapped fluxes by orders of magnitude over a typical time scale of a few days \cite{Horne05JGR, Mourenas23:jgr:upper_limit, Ozeke14, Thorne13:nature}. Flux variations with magnetic local time (MLT) can also be significant, depending on geomagnetic activity \cite{Allison17, Meredith16}.

The balance between electron flux injections and losses depends on a number of factors, such as the level of \blue{substorm activity}, the solar wind speed and dynamic pressure and the local plasma density, which can all modify the strength of the different physical processes at work. Such physical processes also affect the equatorial pitch-angle distribution of electrons and the corresponding flux distribution along magnetic field lines \cite{Kennel&Petschek66, Li13:POES, Mourenas14:fluxes}. Therefore, developing a model of the distribution of omnidirectional electron fluxes along fixed geomagnetic field lines as a function of $L$, MLT, and \blue{substorm activity} can be useful for assessing the dominant physical processes at a given time and location.

Besides, internal charging represents a major hazard for satellites \cite{Chen21:TID, Zheng19}. It is caused by high fluxes of energetic to relativistic ($>200$ keV) electrons. The total dose of such radiation can be used to estimate the charge deposition inside spacecraft electronic components and the probability of electrostatic discharges in dielectrics. Increases in lower energy $\sim1-200$ keV electron fluxes can similarly lead to satellite surface charging, potentially resulting in electrostatic discharges that may damage solar array panels \cite{Zheng19}. Therefore, it is crucial for spacecraft designers to estimate the total radiation dose expected during a satellite lifetime \cite{Zheng19}. Spacecraft operators also need predictive (or probabilistic) models with a capability to forecast periods of particularly elevated electron fluxes, which may allow mitigating the impact of space weather hazards -- for instance, by temporarily shutting down satellite operations. 

While various past models of radiation belt \blue{omnidirectional} electron fluxes \cite<e.g., see>[and references therein]{Boynton16b, Simms23} have focused on geosynchronous orbit (GEO), the total ionizing dose risks for satellites in Low Earth Orbit (LEO), Highly Elliptical Orbit (HEO), and Medium Earth Orbit (MEO), are somewhat less well known. Electron fluxes have been found to vary coherently from LEO to higher altitudes on the same $L$-shell at $L\leq7$ \cite{Kanekal01, Shane23}. Building on this coherency, various machine learning models have recently been developed to nowcast or forecast the \blue{omnidirectional} electron flux at LEO, HEO, MEO, and GEO, based on conjugate measurements of electron fluxes by LEO or Global Positioning System (GPS) spacecraft and several solar wind or geomagnetic indices \cite{Boyd23, Lima20, Smirnov20}. Other recent models provide deterministic or probabilistic forecasts of the \blue{omnidirectional} electron flux at different orbits, based only on past solar wind or geomagnetic indices, the expected future level of such indices, or sequences of time-integrated values of past geomagnetic indices \cite{Boynton2019, Ma22:flux, Mourenas22:jgr:climatology, Simms23}.

In the present work, the full dataset of electron fluxes measured by Electron Loss and Fields Investigation (ELFIN) CubeSats \cite{Angelopoulos20:elfin} at low altitude ($\sim$450 km) in 2020-2022 is used to develop a model of omnidirectional $60-1500$ keV electron fluxes, at altitudes varying from 150 km to 20,000 km along $L$-shells ranging from $L=1.5$ to $L=10$, as a function of past \blue{substorm activity}. \blue{Making use of a novel method, building on adiabatic transport theory and quasi-linear diffusion theory,} we provide a self-consistent model of omnidirectional electron flux, solely based on measurements in low Earth orbit. Problems related to conjunction and intercalibration uncertainties arising when combining data from different spacecraft on different orbits \blue{(as in most previous models)} are naturally avoided, resulting in inferred electron fluxes intrinsically coherent at all altitudes along a given geomagnetic field line. Stepwise regression is used to obtain an analytical model of electron flux based on inferred fluxes. \citeA{Simms23} have shown that electron flux models obtained through stepwise regression procedures can reach roughly similar accuracy and predictive ability as neural network models, while being simpler and more portable. In the present model, average electron fluxes are provided for four different ranges of two different indices of past \blue{substorm activity}, over the preceding 4 hours or over the previous 72 hours. \blue{This contrasts with most previous models, which are usually provided for one fixed set of indices and not in two separate versions for two different indices, as here.}

Hereafter, Section \ref{sec:method} describes the methodology employed to infer omnidirectional electron fluxes at various altitudes along a given geomagnetic field line from ELFIN measurements of pitch-angle resolved electron fluxes at 450 km altitude. The stepwise optimization procedure and the resulting analytical model of omnidirectional electron flux are described in Section \ref{sec:model}. In Section \ref{sec:discussion}, the model is validated by comparisons with other spacecraft measurements near the magnetic equator, and several physical implications of the results are discussed. 

\section{Methods and Data}\label{sec:method}

\subsection{ELFIN Dataset}

In the following, ELFIN CubeSats (referred to respectively as ELFIN-A and ELFIN-B) measurements of $60-1500$ keV precipitating, trapped (or quasi-trapped), and backscattered electron fluxes \cite{Angelopoulos20:elfin} at $400-450$ km altitude are used to develop a novel model of omnidirectional electron fluxes at all altitudes along each geomagnetic field line, down to 150 km altitude below ELFIN and up to $\sim20,000$ km altitude above it, making use of adiabatic transport theory and quasi-linear diffusion theory. The two CubeSats were launched in a nearly polar circular orbit of 90 minutes period. They have provided energy and pitch-angle resolved measurements of electron fluxes during the spacecraft spin period of 2.85 seconds, from 2019 to 2022. The energy resolution is $\Delta E/E \sim 40$\% and the resolution in local pitch-angle $\alpha$ is $\Delta\alpha\sim 22.5^\circ$ from $\alpha=0^\circ$ to $\alpha=360^\circ$. This data set has been used in various recent studies to investigate wave-particle interactions and their effects on electron fluxes and has been described extensively in previous papers \cite<e.g., see>[]{Angelopoulos23:ssr, Mourenas21:elfin, Mourenas23:jgr:upper_limit, Zhang22:natcom}. 

In the present study, the 2020-2022 datasets of ELFIN A and B are used, representing respectively 5,200 and 4,100 separate time periods \cite<each period is one ELFIN orbit or so-called {\it science zone}; see>[]{Tsai24:review} with available data, with a mean duration of 8 min. \blue{\citeA{Ma22:flux} have shown that $50-900$ keV electron fluxes at $L=2.6-6.0$ are better correlated with substorm activity (through $AE$ or $AL$ indices) than with $SYM-H$, solar wind dynamic pressure $Pdyn$, or solar wind speed $V_{sw}$, although additional correlations exist with these other parameters. Note that in the present study, geomagnetic or substorm activity can be taken into account only at $L>3.5$, due to more sparse ELFIN data at lower $L<3.5$. At $L=4-6.6$, several studies have also shown that maximum or average $0.12-2$ MeV electron fluxes are better correlated with maximum or time-integrated $AE$ or $AL$ than with time-integrated $SYM-H$, $Dst$, or $ap$ (related to $Kp$), or minimum $Dst$, or instantaneous $Kp$ \cite{Hua22b, Mourenas19:Impact, Smirnov20}. Substorm-time electron injections at all $L$-shells can also be taken into account using $AE$ or $AL$ indices \cite{Tang16b, Gabrielse19}. Based on these previous results, and for the sake of simplicity, we decided to use in the present work a single parameter, substorm activity, quantified by $AE$. The main goal of present study is indeed to provide a simple, practical analytical flux model, inferred from measurements at low Earth orbit using a new method, and to demonstrate the validity of this novel method. Our new flux model could probably be improved by additionally taking $SYM-H$ or $Pdyn$ into account, but this would make it significantly more complex, and this is left for future work.}

Two different parameters, $AE^\star$ and $AE^{\star\star}$, are used to quantify \blue{substorm activity}. The $AE^\star$ index (in nT) is defined as the mean value of the $AE$ index during the previous 4 hours, allowing to roughly take into account \blue{the time it takes for $\sim10-100$ keV plasma sheet electrons to drift azimuthally from the midnight sector where they are injected to all other MLTs around the Earth \cite{Meredith04, bookSchulz&anzerotti74}, so that these electrons can locally generate whistler-mode waves at all MLTs} and provide a seed electron population that can be accelerated to higher energies. Hereafter, the SuperMAG $\mathit{SME}$ index is employed as a fair proxy for the $AE$ index \cite{Gjerloev12}. The $AE^{\star\star}$ index (in nT$\cdot$hr), defined as the time-integrated $AE$ (or $\mathit{SME}$) during the preceding 72 hours, is used to take into account the peculiar effects of high time-integrated \blue{substorm activity} (i.e., of prolonged injections, ULF wave-driven radial diffusion and electron energization, and chorus wave-driven electron acceleration), which are known to produce the highest omnidirectional electron fluxes in the outer radiation belt \cite{Hua22b, Mourenas19:Impact, Mourenas22:jgr:climatology}. These two different parameters, $AE^\star$ and $AE^{\star\star}$, can therefore be used to separate the effects of rapid physical processes from those of prolonged cumulative processes in electron flux variations.

\subsection{Omnidirectional Electron Flux at Altitudes Lower than ELFIN}\label{sec:adiabatic}
 
Firstly, ELFIN measurements at an altitude $h_0\sim 450$ km of pitch-angle resolved (i.e. directional) differential electron fluxes $J(h_0,\alpha_{h_0},E)$, in units of e/cm$^2$/s/sr/MeV, with $\alpha_{h_0}$ the local pitch-angle, are used to infer omnidirectional fluxes $J_{omni}(h,E)$ at $h\leq h_0$, down to $h=150$ km below the spacecraft. \blue{In this study, we use the approximation of a conserved first adiabatic invariant after averaging over electron gyro-rotation (i.e., the guiding center approximation), and we also use (further below) the usual theoretical formulation of the electron bounce period, under the assumption of a slowly varying background geomagnetic field compared with both the electron gyroperiod and its bounce period \cite{bookSchulz&anzerotti74}. The validity of these approximations was checked numerically, showing that errors remain less than $\sim1$\% for $L\lesssim 6$ and $E<5$ MeV \cite{Soni20, Soni21}, which is largely sufficient for the present purposes.} For equipotential magnetic field lines and slow variations of the near-Earth geomagnetic field compared to an electron gyroperiod, the conservation of the number and energy of electrons and of the magnetic flux and first adiabatic invariant between $h_0$ and $h$ along the same field line leads to the conservation of the gyrotropic unidirectional flux, $J(h,\alpha_h(\alpha_{h_0}),E)=J(h_0,\alpha_{h_0},E)$ \cite{bookRoederer70, bookSchulz&anzerotti74, Walt94}. The omnidirectional differential electron flux (in e/cm$^2$/s/MeV) at $h_0$ is given by
\begin{equation}
J_{omni}(h_0,E) = 2\pi \, \int_{0}^{\,\pi} J(h_0,\alpha_{h_0},E)\sin\alpha_{h_0} \, d\alpha_{h_0}.
\label{eq1}
\end{equation}
For the sake of simplicity, we hereafter use an eccentric (off-centered) and inclined dipolar external geomagnetic field model \cite<e.g., see>[]{Koochak17}, which is a reasonable approximation to the actual geomagnetic field for $h\in[150; 20,000]$ km, together with the International Geomagnetic Reference Field \cite<IGRF; see>[]{Thebault15} magnetic latitude of the spacecraft provided in the ELFIN data set. The downward part of the omnidirectional flux, $J_{omni}^{down}$, at an altitude $h<h_0$ on the same field line as ELFIN can then be written as
\begin{equation}
J_{omni}^{down}(h,E) = 2\pi \, \int_{0}^{\alpha_{max,0}} J(h,\alpha_h(\alpha_{h_0}),E) \, \sin\alpha_h(\alpha_{h_0})\, \frac{{ \partial\alpha_h }}{{ \partial\alpha_{h_0} }} \, d\alpha_{h_0},
\label{eq2}
\end{equation}
where $\alpha_{max,0}$, the local pitch-angle at $h_0$, corresponds to a local pitch-angle $\alpha_h=\pi/2$ at $h$ \cite{Ni2009}. Conservation of the first adiabatic invariant yields $\sin^2\alpha_h/\sin^2\alpha_{h_0} = B(h)/B(h_0)$, with $B(h)$ the geomagnetic field strength \cite{bookRoederer70, bookSchulz&anzerotti74}, giving $\sin\alpha_{h_0}\leq \sin\alpha_h$ for $h\leq h_0$. Therefore, $J_{omni}(h,E)$ at $h\leq h_0$ is fully determined by $J(h_0,\alpha_{h_0},E)$ at $h_0$ together with conservation of unidirectional flux, energy, and first adiabatic invariant. Since $\partial\alpha_h/\partial\alpha_{h_0}= B(h)\sin\alpha_{h_0}\cos\alpha_{h_0}/(B(h_0)\sin\alpha_h\cos\alpha_h)$, this finally gives:
\begin{equation}
J_{omni}^{down}(h,E) = 2\pi \, \frac{{ B(h) }}{{ B(h_0) }} \, \int_{0}^{\alpha_{max,0}} J(h_0,\alpha_{h_0},E) \frac{{ \sin\alpha_{h_0}\cos\alpha_{h_0} }}{{ \sqrt{1 - \frac{{ B(h) }}{{ B(h_0) }} \sin^2\alpha_{h_0} } }}\,  \, d\alpha_{h_0},
\label{eq3}
\end{equation}
where $J(h_0,\alpha_{h_0},E)$ is the directional differential flux inferred, by cubic spline interpolation, from the fluxes measured by ELFIN on different pitch-angle intervals ($B(h)/B(h_0)$ is evaluated for an inclined eccentric dipolar geomagnetic field). The integral in Equation (\ref{eq3}) is calculated using the QUAD package from the Scientific Python (SciPy) library, which allows to smoothly take into account a singularity at $\alpha_h=\pi/2$. The upward omnidirectional flux $J_{omni}^{up}(h,E)$ is similarly obtained from the (upward) directional flux at $\alpha_{h_0}\in[\pi/2,\pi]$ measured by ELFIN, and the total omnidirectional electron flux is $J_{omni}=J_{omni}^{down}+J_{omni}^{up}$. 

\subsection{Omnidirectional Electron Flux at Altitudes Higher than ELFIN}

In several previous works, equatorial electron fluxes at high altitudes have been inferred from low-altitude flux measurements based on the observed high correlations between conjugate low-altitude and high-altitude fluxes, or else by using statistical pitch-angle distributions from the Van Allen Probes \cite{Allison18, Lima20, Boyd23}. In the present study, we adopt a different approach. At altitudes $h\geq h_0$, the part of the omnidirectional flux at $\alpha_h\in[0, \alpha_{hAL}]$ and at $\alpha_h\in[\pi-\alpha_{hAL}, \pi]$, with $\alpha_{hAL}$ the adiabatic limit (AL) at $h$ corresponding to $\alpha_{h_0}=\pi/2$ on the same field line, can still be directly inferred from ELFIN measurements using adiabatic transport theory, as in Section \ref{sec:adiabatic}. However, this part of $J_{omni}(h,E)$ becomes smaller and smaller at higher altitudes because $\alpha_{hAL}$ decreases as $h$ increases above $h_0$. 

We assume that a quasi-equilibrium pitch-angle electron distribution has been reached after hours to days of wave-particle interactions. \blue{This requires a roughly steady level of MLT-averaged wave power over the several hours (for high $AE^\star$ or $AE^{\star\star}$) to $24$ hours (for low $AE^\star$ or $AE^{\star\star}$) preceding ELFIN measurements on a given $L$-shell. Previous works have shown that this condition is usually satisfied during quiet and moderately disturbed periods \cite{Mourenas21:elfin,Shane23}. Statistical wave models further indicate that the wave power is well correlated with the $AE$ level, implying that periods of high $AE^\star$ or $AE^{\star\star}$ (corresponding to 4-hour to 3-day periods of high $AE$ preceding ELFIN measurements) should also correspond in general to a high and roughly steady level of MLT-averaged wave power over the several hours to days preceding ELFIN measurements \cite{Agapitov19:fpe, Meredith07}. In this case, we can} use quasi-linear diffusion theory \cite{Kennel&Petschek66, Li13:POES} to infer high-altitude fluxes at local pitch-angles $\alpha_h\in[\alpha_{hAL}, \pi-\alpha_{hAL}]$ from low-altitude fluxes measured by ELFIN, \blue{on the same geomagnetic field line}. This is achieved with the help of simultaneous ELFIN measurements of the net precipitating to trapped flux ratio, $J_{prec}/J_{trap}$. As in previous works, the net precipitating flux, $J_{prec}$, directly precipitated by wave-particle interactions, is defined as the measured precipitating flux averaged inside the local bounce loss cone, minus the average upward flux $J_{up}$ backscattered inside the same bounce loss cone, where $J_{up}$ is used as a proxy for the flux backscattered by the atmosphere from the opposite hemisphere on the same field line over times long compared to a bounce period \cite{Mourenas21:elfin, Mourenas23:jgr:upper_limit}. This estimate of the net precipitating to trapped flux ratio $J_{prec}/J_{trap}$ relies on the assumption that a majority of backscattered electrons should remain within the same energy bin of width $\Delta E/E\approx40$\% \cite<in agreement with simulations, see>[]{MarshallBortnik18, Selesnick04} and also assumes a symmetric system about the magnetic equator. But since random errors should partly cancel out after averaging the inferred high altitude fluxes over many measurements at various locations, the estimated time-averaged high altitude flux should remain approximately correct in the presence of small deviations from symmetry.

At $L\sim1.5-10$ in the inner magnetosphere and near-Earth plasma sheet, in the presence of typical populations of incoherent whistler-mode waves or of mostly short and intense whistler-mode wave packets with random frequency and phase jumps \cite{Gao22:Luphi_whistlers, He21, Mourenas22:jgr:ELFIN, Zhang20:grl:phase, Zhang20:grl:frequency}, the quasi-linear diffusion theory is expected to remain approximately valid \cite{Artemyev22:jgr:NL&QL, Mourenas21:elfin, Mourenas22:jgr:ELFIN, Zhang20:grl:phase}. Quasi-linear theory probably also holds in the presence of other types of waves with similar characteristics, such as EMIC waves \cite{Angelopoulos23:ssr, Remya17}.

Quasi-linear diffusion theory \cite{Kennel&Petschek66} provides an approximate relationship between the effective pitch-angle diffusion rate $D_{\alpha\alpha}$ of electrons at the loss cone angle and the average net precipitating to trapped flux ratio $J_{prec}/J_{trap}$ measured at ELFIN CubeSats \cite{Mourenas23:jgr:upper_limit, Mourenas24:jgr:ELFIN&KPlimit}:

\begin{equation}
z_0 \,\, \simeq \,\, \left(10^4 + 260\, \frac{{ J_{trap} }}{{ J_{prec} }} \right)^{1/2} \, -100,
\label{eq4}
\end{equation}
with a moderate error $<25$\% for $J_{prec}/J_{trap}<0.85$, $z_0 = 2\alpha_{eq,LC}/(D_{\alpha\alpha} \tau_B)^{1/2}$, $\tau_B(E,L)$ the electron bounce period, and where $D_{\alpha\alpha}(E,L)$ is calculated at the equatorial loss cone angle $\alpha_{eq,LC}$, which corresponds to a local $\alpha=\pi/2$ at $h=100$ km \cite{bookSchulz&anzerotti74}. The average quasi-equilibrium directional electron flux at $\alpha_h>\alpha_{hAL}$ is given by quasi-linear theory \cite{Kennel&Petschek66, Mourenas24:jgr:ELFIN&KPlimit}:

\begin{equation}
\frac{{ J(h,\alpha_h,E) }}{{ J(h,\alpha_{hAL},E) }} \, \approx \, \frac{{ 1 + z_0 \frac{{ I_1(z_0) }}{{ I_0(z_0) }} \ln\left(\frac{{ \sin\alpha_{eq,h} }}{{ \sin\alpha_{eq,LC} }} \right) }}{{ 1 + z_0 \frac{{ I_1(z_0) }}{{ I_0(z_0) }} \ln\left(\frac{{ \sin\alpha_{eq,hAL} }}{{ \sin\alpha_{eq,LC} }} \right) }},
\label{eq5}
\end{equation}
where $I_x$ is the modified Bessel function of the first kind, and $\alpha_{eq,h}$ and $\alpha_{eq,hAL}$ are the equatorial pitch-angles corresponding to $\alpha_h$ and $\alpha_{hAL}$, respectively. Using Equations (\ref{eq4}) and (\ref{eq5}), the directional electron flux at $\alpha_h\in[\alpha_{hAL}, \pi/2]$ can be inferred from ELFIN measurements of $J_{prec}/J_{trap}$ and $J(h_0,\alpha_{h_0}=\pi/2,E) = J(h,\alpha_{hAL},E)$. We also assume that $J(h,\pi-\alpha_h,E)\simeq J(h,\alpha_h,E)$ to first order for $\alpha_h\in[\pi/2, \pi-\alpha_{hAL}]$. This assumption is justified for a roughly symmetric system about the magnetic equator, especially for $J_{omni}$ since integrating over all pitch-angles reduces the average relative error. The total omnidirectional flux $J_{omni}(h,E)$ is finally obtained by summing two parts: a first, adiabatic part at $\alpha_h<\alpha_{hAL}$ and $\alpha_h>\pi-\alpha_{hAL}$ calculated as in Section \ref{sec:adiabatic}, and a second part calculated over the remaining $\alpha_h$ range by integrating as in Equation (\ref{eq1}) the directional fluxes obtained from Equation (\ref{eq5}).

Note that the above method is valid only when $J_{prec}/J_{trap}<0.85$, that is, in a regime of weak diffusion. When $J_{prec}/J_{trap}>0.85$, we enter a regime of strong diffusion, where $z_0\simeq1$ \cite{Kennel69}. Equation (\ref{eq5}) indicates that for $z_0=1$, $J(h,\alpha_h,E)$ increases only very weakly as $\alpha_h$ increases. In the strong diffusion regime, the actual $D_{\alpha\alpha}$ can even exceed the theoretical level corresponding to $z_0=1$, leading to a constant $J(h,\alpha_h,E)$ at $\alpha_h>\alpha_{hAL}$. When $J_{prec}/J_{trap}>0.85$, it is therefore reasonable to use the simple approximation $J(h,\alpha_h,E) \approx J(h,\alpha_{hAL},E)$ for $\alpha_h>\alpha_{hAL}$, with a corresponding error on $J(h,\alpha_h,E)$ usually much smaller than a factor of 2.

\subsection{Expected Validity Domain}

Some limitations of the present method for inferring fluxes at $h>h_0$ are worth mentioning. Equation (\ref{eq5}) has been derived by assuming, as in the original work by \citeA{Kennel&Petschek66}, that $D_{\alpha\alpha}$ is varying with $\alpha_{eq}$ roughly like $\approx1/\cos\alpha_{eq}$ at $\alpha_{eq}<80^\circ-90^\circ$. Analytical estimates, validated by numerical simulations, have shown that for quasi-parallel whistler-mode waves, the actual variation of $D_{\alpha\alpha}$ with $\alpha_{eq}$ is usually closer to $\approx1/\cos^2\alpha_{eq}$ above $\sim100$ keV \cite{Agapitov18:jgr, Artemyev13:angeo, Li15, Mourenas12:JGR}. However, depending on wave power and frequency distributions and plasma density, the variation of $D_{\alpha\alpha}$ may sometimes become similar to $\approx\cos\alpha_{eq}$, especially at $L<3.5$ and low energy \cite{Green20, Li15, Ma17:vlf, Ma22:VLF}. For these two alternative variations of $D_{\alpha\alpha}$ with $\alpha_{eq}$, all terms of the form $\ln(\sin\alpha_{eq})$ in Equation (\ref{eq5}) have to be replaced, in the first case by $\cos\alpha_{eq} + \ln(\tan(\alpha_{eq}/2))$, and in the second case by $\ln(\tan\alpha_{eq})$. In a dipolar geomagnetic field at $L>1.5$, $J_{omni}(h,E)$ values inferred using Equation (\ref{eq5}) at $\alpha_{eq}<80^\circ-85^\circ$ \cite<assuming the same variation of $D_{\alpha\alpha}$ with $\alpha_{eq}$ as>[]{Kennel&Petschek66} remain within a factor of $\approx1.5-2$ from $J_{omni}(h,E)$ values inferred using the above-discussed two alternative variations of $D_{\alpha\alpha}$ with $\alpha_{eq}$, indicating the reliability of Equation (\ref{eq5}).

However, the eccentric dipole approximation to the actual geomagnetic field remains reasonable only up to $h\approx20,000$ km during disturbed periods \blue{\cite<e.g., see>[]{Berube06, Ganushkina02, Roederer18}.} 
This suggests that the accuracy of the $J_{omni}(h,E)$ model should be ensured only for
\begin{equation}
h \,<\, h_{max} \approx 20,000 \,\, {\rm km}.
\label{eq6}
\end{equation}
\blue{This means that the present $J_{omni}(h,E)$ model should remain approximately valid at the magnetic equator only up to $L\approx4.2$, whereas at $L>4.5$ it should remain approximately valid only sufficiently far from the magnetic equator, corresponding to altitudes $h<h_{max}\approx20,000$ km. Note also that the present model is valid only on closed magnetic field lines, where electrons remain stably trapped. Based on numerical calculations of the last closed magnetic field lines \cite{Olifer18, Roederer18}, the model should remain valid at all MLTs up to $L\simeq9-10$ when $Kp\leq4$, a condition roughly equivalent to $AE\leq700$ nT \cite{Rostoker91}, while during strong geomagnetic storms and substorms with $Kp>4$ the model should still remain valid at all MLTs up to at least $L\simeq6-7$ at times when $Dst>-100$ nT. In 2020-2022, $Dst$ always remained higher than $-100$ nT, except for one storm with a minimum $Dst$ of $-105$ nT.} 

The maximum altitude $h_{max}$ corresponds to maximum equatorial pitch-angles $\alpha_{eq,max}\sim\sin^{-1}\left(\left(1 + h_{max}/R_E\right)^{3/2}/\left(L^{3/2}(4 - 3(1 + h_{max}/R_E)/L\right)^{1/4})\right)$ for the applicability of the present method, giving $\alpha_{eq,max}\lesssim70^\circ$, $45^\circ$, and $25^\circ$ at $L>4$, $5$ and $6$, respectively. 

In addition, the variation of $D_{\alpha\alpha}$ with $\alpha_{eq}$ can sometimes be more complex than the above-discussed simple scaling laws. But taking into account all waves (and Coulomb collisions) at $L=1.5-6$ within the plasmasphere or in a plasmaspheric plume, $D_{\alpha\alpha}$ should usually not decrease by much more than a factor of $\sim\tan\alpha_{eq,LC}/\tan\alpha_{eq}$ as $\alpha_{eq}$ increases from $\alpha_{eq,LC}$ to $\alpha_{eq,max}=\alpha_{eq}(h_{max})$ for $0.1-1.5$ MeV electrons \cite{Angelopoulos23:ssr, Green20, Li15, Ma17:vlf, Ma22:VLF, Shi24:emics, Wong22}. Then, the inferred $J_{omni}(h,E)$ should remain within a factor of $\approx2$ from the actual $J_{omni}(h,E)$ at $h<h_{max}$. 

Outside the plasmasphere, chorus wave-driven energy diffusion can compete with pitch-angle diffusion \cite{Horne05JGR, Summers98}, but Van Allen Probes observations show that this should not significantly modify the increase of $J(\alpha_{eq})$ with $\alpha_{eq}$ up to at least $\alpha_{eq}\approx50^\circ$ for $E<1.5$ MeV  \cite{Li14:storm}. At $L\sim 6-10$, magnetic field line curvature scattering \cite{Young02} still leads to an increase of $D_{\alpha\alpha}$ up to $\alpha_{eq,max}=\alpha_{eq}(h_{max})$ \cite{Artemyev13:angeo:scattering}, and drift shell splitting should not strongly modify $J(\alpha_{eq})$ below $\alpha_{eq}(h_{max})$ \cite{Selesnick02}, which should keep the inferred $J_{omni}(h,E)$ within a factor of $\sim1.5$ from the actual $J_{omni}(h,E)$ at $h<h_{max}$. 

Finally, it is worth emphasizing that our model of omnidirectional electron flux is based on time-averaged inferred fluxes $J_{omni}$, averaged inside each parameter bin over at least 25 (and often much more) ELFIN measurements performed at different times. Random errors on individual inferred $J_{omni}$ values will partly cancel each other out. Therefore, the average inferred $J_{omni}(h<h_{max},E)$ is expected to remain less than a factor of $\sim1.5$ (at $L>3.5$) to $\sim2$ (at $L<3.5$) from the actual average $J_{omni}(h,E)$. \blue{The error is expected to be largest when the assumption that $D_{\alpha\alpha}$ does not decrease by more than a factor of $\sim\tan\alpha_{eq,LC}/\tan\alpha_{eq}$ as $\alpha_{eq}$ increases from $\alpha_{eq,LC}$ to $\alpha_{eq}(h_{max})$ is not verified, which should mainly occur at $L<3.5$ for low energy electrons.}

\section{Model of Omnidirectional Electron Fluxes}\label{sec:model}

\subsection{Data Selection}

The electron detector onboard ELFIN Cubesats provides differential electron fluxes measured in 16 logarithmically spaced energy channels (each with a full width of $\Delta E/E\sim 40$\%) whose central values extend from $60$ keV to 6.5 MeV \cite{Angelopoulos20:elfin}. Over a spacecraft spin period of 2.85 s, an ELFIN CubeSat provides two complete electron flux measurements of the entire $180^\circ$ local pitch-angle distribution, with a $\sim22.5^\circ$ resolution, resolving quasi-trapped, precipitating, and upward-moving electrons backscattered by the atmosphere \cite{Angelopoulos20:elfin}.

Before computing $J_{omni}$, a strict data screening procedure is used in order to only keep the most reliable electron flux data:
\begin{itemize}
    \item firstly, if $J(\alpha_{h_0})<100$ e/cm$^2$/s/sr/MeV at a given pitch-angle, or if the associated number of counts per second is below 5 for a given channel, the measured flux is considered to be null for this channel, in order to only keep fluxes above instrument noise level \cite{Mourenas24:jgr:ELFIN&KPlimit}. This conservative approach should only lead to a very slight underestimation of the final time-averaged omnidirectional flux, since such cases correspond to very low to null fluxes, much smaller than retained fluxes,
    \item at a given time, for a given energy channel, at least three pitch-angle bins must be associated with non-zero fluxes,
    \item if there are exactly three pitch-angle bins associated with non-zero fluxes, they must be adjacent pitch-angles (to exclude fluxes with abnormal fluctuations),
    \item the flux measured at the first pitch-angle just above the loss cone angle must be non-zero (to have a non-null quasi-trapped flux) and higher than the flux measured just below \cite<opposite cases may correspond to occasional rapid fluctuations or to isolated bursts of very oblique waves leading to a fully nonlinear electron transport that cannot be described by quasi-linear theory, see>[]{Zhang22:natcom}.
\end{itemize}

We then compute, for acceptable measurements, the time-averaged corresponding values of $J_{prec}$ and $J_{trap}$, discarding cases for which one of these values turns out to be non-positive or not calculable. As a result, about 20\% of the full 2020-2022 ELFIN data set have been retained, the overwhelming majority of data rejections being due to the presence of less than three pitch-angles with non-zero flux at a given time and at a given energy.

Finally, the omnidirectional fluxes $J_{omni}(h,E,L)$ are inferred from ELFIN data at 18 pre-determined altitudes between $h=150$ km and $h=20,000$ km (at 150, 200, 250, 350, 450, 600, 800, 1000, 1200, 1600, 2000, 4000, 6000, 8000, 11000, 14000, 17000, 20000 km), with a shorter step at lower altitudes where flux variations are stronger, and for $E\in[0.06,1.5]$ MeV and $L\in[1.5,10]$ using the methods described in Section \ref{sec:method}. 
It is worth noticing that the values of $J_{prec}$ and $J_{trap}$ used to compute the net precipitating to trapped flux ratio (necessary to establish the weak diffusion condition, and then considered in Equation \ref{eq4} and, for $J_{trap}$, in the denominator of the left-hand side of Equation \ref{eq5}) are averaged on all positive available values over a 18-second sliding window, to provide more reliable fluxes, time-averaged over a period much longer than a bounce period \cite{Mourenas21:elfin}, also mitigating possible time-aliasing effects \cite{Angelopoulos23:ssr,Zhang22:microbursts}.

We use ELFIN A data as the training subset, and ELFIN B data as a validation subset. For each subset, omnidirectional fluxes are averaged inside each parameter bin $(AE^\star,E,L,h)$ or $(AE^{\star\star},E,L,h)$. To do so, $L$ is rounded to the nearest quarter of an integer, and $AE^\star$ and $AE^{\star\star}$ values are each associated to one of the four following levels of instantaneous or time-integrated \blue{substorm activity}:
\begin{itemize}
    \item quiet ($AE_0^\star$ and $AE_0^{\star\star}$ levels): $AE^\star < 100$ nT or $AE^{\star\star} < 10^4$ nT$\cdot$h
    \item moderate ($AE_1^\star$ and $AE_1^{\star\star}$ levels): $100$ nT $< AE^\star < 300$ nT or $10^4$ nT$\cdot$h $ < AE^{\star\star} < 2\cdot10^4$ nT$\cdot$h
    \item active ($AE_2^\star$ and $AE_2^{\star\star}$ levels): $300$ nT $< AE^\star < 500$ nT or $2\cdot10^4$ nT$\cdot$h $ < AE^{\star\star} < 3\cdot10^4$ nT$\cdot$h
    \item very active ($AE_3^\star$ and $AE_3^{\star\star}$ levels): $AE^\star > 500$ nT or $AE^{\star\star} > 3\cdot10^4$ nT$\cdot$h.
\end{itemize}

We obtain that way, for each subset, a time-averaged profile of the flux as a function of altitude, $J_{omni}(h)$, for each $(E,L,AE_i^\star)$ and $(E,L,AE_i^{\star\star})$, where $h$ is varying from 150 km up to a maximum altitude that depends on $L$-shell and does not exceed 20,000 km.

We finally perform a last sorting, by:
\begin{itemize}
    \item discarding values of $J_{omni}(h)$ averaged over less than 25 instantaneous values (each averaged value of $J_{omni}(h)$ considered thereafter will that way be temporally averaged over at least 36 s, since ELFIN gives two measurements of the full $180^\circ$ pitch-angle domain per spin of 2.85 s),
    \item next, deleting whole $J_{omni}(h)$ profiles in altitude with less than four values of $J_{omni}(h)$ at $h>h_0$ (which concretely imposes, given the set of altitudes considered, a maximum altitude $h\geq1,200$ km for an averaged $J_{omni}(h)$ profile to be taken into account),
    \item deleting the few $J_{omni}(h)$ profiles for which $J_{omni}(h)$ decreases by more than $5 \%$ just above $h_0$ (probably due to a drop in the number of $J_{omni}(h)$ values taken into account in the calculation of the average, since $J_{omni}(h)$ is sometimes available only at $h\leq h_0$ due to the impossibility of applying the weak or strong diffusion approximations mentioned above),
    \item deleting $J_{omni}(h)$ profiles associated to an energy higher than 1.5 MeV (such profiles are rare and the corresponding average $J_{omni}$ values are calculated based on only few values),
    \item deleting the small number of average inferred $J_{omni}(E,L,h)$ profiles with values lower than $300$ e/cm$^2$/s/MeV at $h=450$ km, because they are much lower than all the others and mostly correspond to noise. 
\end{itemize}

Note that ELFIN regular {\it science zones} \cite{Tsai24:review} mostly cover $L\sim3-12$, whereas data from lower $L$-shells are much more sparse. Therefore, all the average inferred $J_{omni}(E,L,h)$ profiles at $L\in[1.5,3.5]$, much rarer than at $L\in[3.5,10]$, are regrouped in one set independently of the $AE$ level. This procedure, necessary to obtain reliable values for all $(E,L)$ pairs, is justified by the weaker variation of time-averaged $0.06-1.5$ MeV electron fluxes with geomagnetic activity at $L\leq3$ than at $L>3.5$ \cite<e.g., see>[]{Mourenas17, Reeves16}. As a result, however, the model includes no dependence on \blue{substorm activity} at $L\in[1.5,3.5]$.

We finally get, for each \blue{substorm activity} indicator, a training data set, derived from ELFIN A measurements, and a validation data set, derived from ELFIN B measurements, consisting each of 1200 $J_{omni}(h)$ profiles for both $AE^\star$ and $AE^{\star\star}$.

\subsection{Multivariate Optimization Analysis of Electron Flux Variations}

In this section, multivariate optimization analysis \cite{Gill2020} is used for specifying a model of omnidirectional electron flux $J_{omni}(h,E,L)$ as a function of altitude ($h$), energy ($E$), and $L$, for the four aforementioned levels of preceding \blue{substorm activity} (defined by $AE^{\star}$ or $AE^{\star\star}$ parameters), based on $J_{omni}$ values inferred from ELFIN electron flux data collected in 2020-2022. We first examine $J_{omni}(h,E,L)$ values averaged over MLT, to obtain a much larger number of data points in each parameter bin, and also because past studies have shown that the variations of electron flux with MLT usually remain moderate \cite<a factor of $\approx2$ between dawn and dusk for $AE^\star<1000$ nT, see>[]{Allison17, Meredith16}. The electron equatorial pitch-angle distribution, formed by wave-particle interactions over many azimuthal drift periods \cite{bookSchulz&anzerotti74}, also remains roughly similar at all MLTs at $L\sim1.5-6$ during not-too-disturbed periods \cite{Shi16}. The MLT variations of $J_{omni}$ will nevertheless be examined further below. 

The three independent variables $h$, $E$, and $L$, as well as the additional independent parameters $AE^{\star}$ or $AE^{\star\star}$, are selected here, because it is well known that electron fluxes vary with altitude, electron energy, $L$-shell and \blue{substorm activity}. The magnetic latitude (MLAT) could have been used as an alternative to $h$ since $h$ and MLAT are directly related in a dipolar field but, as our goal is to provide model fluxes at given altitudes, directly using $h$ is more practical in this case. 

The variation of the omnidirectional electron flux with altitude is taken into account by adopting the functional form
\begin{equation}
J_{omni}(h) = B \cdot \left(\ln(h+200)\right)^C,
\label{eq7}
\end{equation}
where $J_{omni}$ is hereafter in units of e/cm$^2$/s/MeV, $h$ in km, and the two variables $B$ and $C$ are supposed to depend on $AE$, $L$ and $E$. Four reasons led us to adopt the functional form given in Equation (\ref{eq7}):
\begin{itemize}
    \item initial visual inspection has shown that $J_{omni}(h)$ increases slower at higher $h$;
    \item Equation (\ref{eq5}) implies that $J(h>h_0,\alpha_h=\pi/2)$ depends on a logarithmic function of $h$, $\ln(\sin\alpha_{eq}(\alpha_h=\pi/2)/\sin\alpha_{eq,LC}) \approx \ln(1+3h/2R_E)$;
    \item a simple functional form is usually preferable to avoid overfitting;
    \item the various other functional forms which have been tested, like power-law functions, provided less accurate fits to the data.
\end{itemize}   

The dependence on \blue{substorm activity} (and its impact on flux variations with $E$ and $L$) is taken into account by determining different values of $B$ and $C$ for each of the four levels $AE_i^\star$ and for each of the four levels $AE_i^{\star\star}$ defined above.

After numerous trials, we decided to keep $B$ constant over each separate domain of \blue{substorm activity}. Next, it is important to choose an appropriate functional form for $C(L,E)$ on the basis of theoretical and observational knowledge. First, Equation (\ref{eq7}) implies that $\ln(J_{omni}(h)/(J_{omni}(h_0))$ is proportional to $C(L,E)$. Equation (\ref{eq5}) shows that $\ln(J(h,\alpha_h=\pi/2)/J(h_0,\alpha_{h_0}=\pi/2))$ increases monotonically with $\ln(z_0)$, but slower than $\ln(z_0)$ at large $z_0$, with $z_0= 2\alpha_{eq,LC}(D_{\alpha\alpha}\tau_B)^{-1/2}$ and $D_{\alpha\alpha}$ the pitch-angle diffusion rate at $\alpha_{eq,LC}$. Therefore, $C(L,E)$ should increase monotonically with $\sim|\ln(D_{\alpha\alpha}\gamma/(\gamma^2-1)^{1/2})|/2-2\ln(L)$, although more slowly. The variation of $D_{\alpha\alpha}$ as a function of $L\in[1.5,10]$ and $E\in[0.06,1.5]$ MeV has been provided based on statistics of whistler-mode waves \cite{Agapitov18:jgr, Green20, Ma16:hiss, Ma17:vlf, Ma22:VLF}. It shows two different variations of $D_{\alpha\alpha}$ with $E$, increasing toward higher $E$ at $L<2.5-3$ and decreasing toward higher $E$ at $L>3.5$ \cite<except within the plasma sheet above 0.5-1 MeV near 0 MLT, see>[]{Artemyev13:angeo:scattering}. Second, Equation (\ref{eq7}) indicates that $\ln(J_{omni})$ is proportional to $C(L,E)$, and spacecraft observations have shown the presence of two $L$-shell domains with distinct energy spectra $J_{omni}(E)$, at $L\in[1.5,3.5]$ where the flux rapidly decreases as $E$ increases over $0.1-1.5$ MeV, and the outer belt and near-Earth plasma sheet at $L\approx4-7$ where the flux can decrease sensibly less rapidly toward higher $E$ \cite{Reeves16}. The above theoretical and observational facts therefore suggest using two slightly different forms of $C(L,E)$ in two separate $L$-shell domains: $L\in[1.5,3.5]$ (domain 0) and $L\in[3.5,10]$ (domain 1).

After various trials, the selected functional form of $C(L,E)$ is:
%\begin{eqnarray}
%C = C_0 + C_1 \cdot L + C_2 \cdot L^2 + C_3 \cdot L^3 + \frac{C_4}{L} + \frac{C_5}{L^2} + C_6 \cdot \ln(E)^{C_7} \\ 
%+ \max(0,3.5-L) \cdot (C_8 \cdot \ln(E)^2 + C_9 \cdot \ln(E)^3), \nonumber \label{eq8}
%\end{eqnarray}
\begin{equation}
C = C_0 + C(L) + C(E,L) \label{eq8}
\end{equation}
where:
\begin{equation}
    C(L) = C_1 \cdot L + C_2 \cdot L^2 + C_3 \cdot L^3 + \frac{C_4}{L} + \frac{C_5}{L^2} \nonumber
\end{equation}
and
\begin{equation}
    C(E,L) = 
    \begin{cases}
      C_6 \cdot \ln(E) + (3.5-L) \cdot (C_8 \cdot \ln(E)^2 + C_9 \cdot \ln(E)^3) & \text{if $1.5 < L < 3.5$}\\
      L \cdot (C_6\sqrt{\ln(E)} + \frac{C_7}{\ln(E)} + C_8 \cdot \ln(E)^2 + C_9 \cdot \ln(E)) & \text{if $3.5 < L < 10$}\\
    \end{cases} \nonumber
\end{equation}
with $E$ in keV and $L$ the McIlwain magnetic shell parameter.

For each $(AE_i^\star,E,L)$ and $(AE_i^{\star\star},E,L)$, we use Powell's method \cite<e.g., see>[]{Gill2020} on averaged $J_{omni}(h)$ values to determine the pair $(B,C)$ minimizing the loss function MEF$\times EF_{90}$, where MEF $= \exp(M(|\ln(Q_j)|))$ is the Median Error Factor between model values and actual values of $J_{omni}$ (where $M$ denotes the median and $Q_j$ values are ratios of model to actual values), and $EF_{90}$ is the $90^{th}$ percentile of the Error Factor, $EF=\exp(|\ln(Q_j)|)$. The MEF is equivalent to the Median Symmetric Accuracy introduced by \citeA{Morley18}: it is particularly relevant for electron fluxes varying by orders of magnitude and robust to the presence of outliers and bad data \cite{Morley18, Zheng19}. Its meaning is simply that $50$\% of model fluxes are less than a factor of MEF from actual fluxes. The advantage of using the loss function MEF$\times EF_{90}$ is to simultaneously minimize MEF and $EF_{90}$, thereby ensuring that both $50$\% and $90$\% of model fluxes remain as close as possible to actual fluxes. This should provide a full distribution of model fluxes close to the full distribution of actual fluxes.

Accordingly, for each $(AE_i^\star,E,L)$ or $(AE_i^{\star\star},E,L)$ corresponding to an averaged profile $J_{omni}(h)$ in altitude, we first use Powell's method to find an optimal value of $B$ and $C$. Next, for each $AE_i^\star$ or $AE_i^{\star\star}$ at $L>3.5$ and for all data at $L<3.5$, we determine the coefficient $B$ which allows to best approach these initial $B$ values, using the classical least squares loss function. We then repeat, for each $(AE_i^\star,E,L)$ and $(AE_i^{\star\star},E,L)$, the Powell's optimization over $h$, using now the value of $B$ corresponding to the relevant domain of $AE_i^\star$ or $AE_i^{\star\star}$. This gives us new optimal values of $C$, which we use to get coefficients $C_m$ in Equation (\ref{eq8}) for each $AE_i^\star$ and $AE_i^{\star\star}$ level at $L>3.5$ and for all $AE$ at $L<3.5$. The values of model coefficients $B$ and $C_m$ in each parameter domain are provided in the Appendix. Note that the model is trained over energies $E\in[60,1500]$ keV at $L\in[1.5,10]$ and should not be used outside of these limits. Finally, to obtain a smooth model of $J_{omni}$ through the frontier at $L=3.5$ between the two $L$ domains, $J_{omni}(E,L)$ is interpolated between $L=3.25$ and $L=3.6$.

\begin{figure}
\centering
\includegraphics[width=1\textwidth]{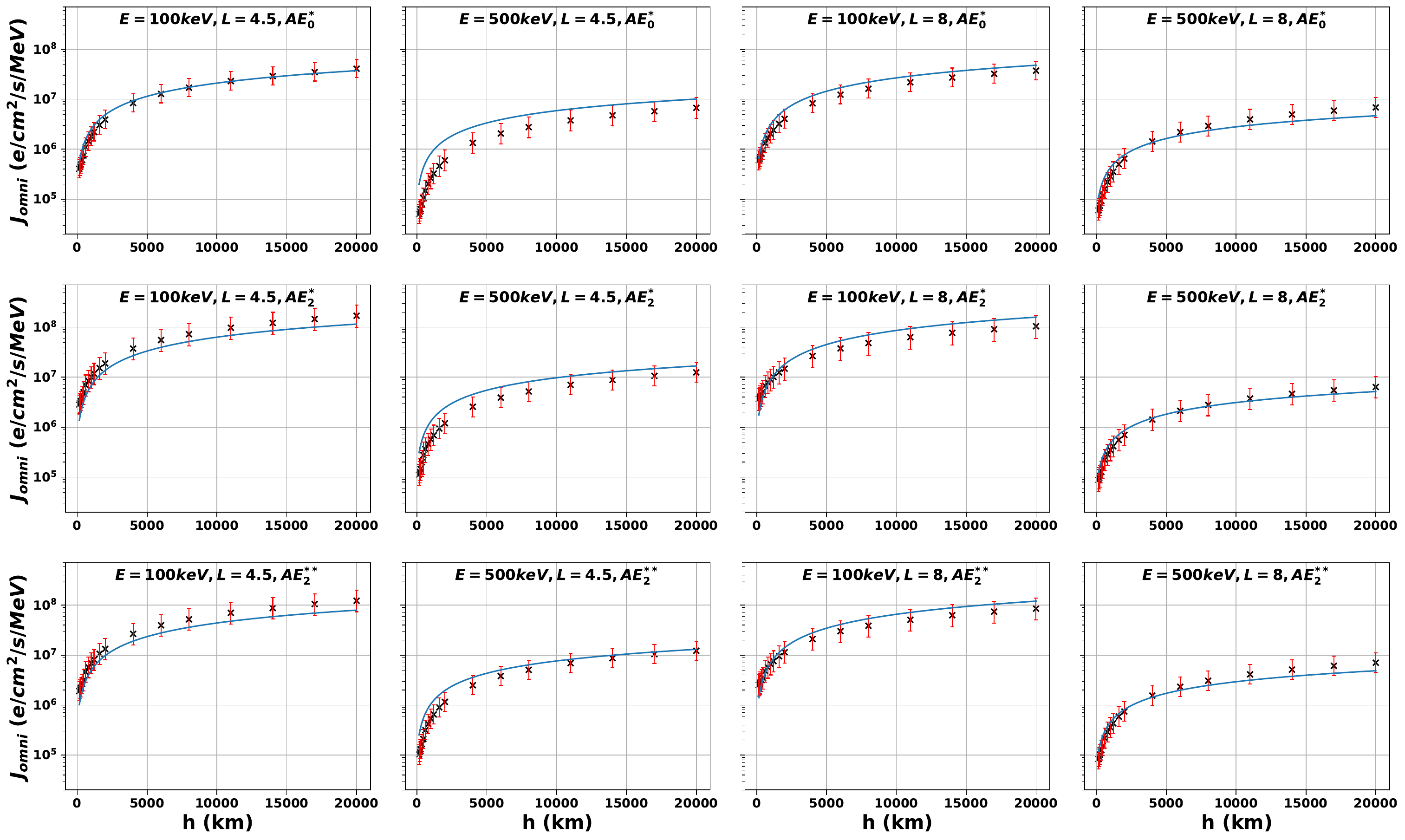}
\caption{Average omnidirectional electron flux $J_{omni}$ as a function of altitude $h$ for different energies, $L$-shells, and \blue{substorm activity} levels, directly inferred from ELFIN A measurements (black crosses, with error bars in red) and given by the model (solid blue line).}
\label{fig1}
\end{figure}

Figure \ref{fig1} shows various examples of average omnidirectional differential electron flux $J_{omni}(h)$ profiles in altitude between $h=150$ km and $h=20,000$ km, either directly inferred from ELFIN A measurements (black crosses) or given by the model (blue solid line), at 100 keV and 500 keV and for different $L$-shells and \blue{substorm activity} levels. The uncertainty of average inferred fluxes is the sum of the uncertainty inherent to the method (estimated as a factor of $\sim1.5$ at $L>3.5$ in Section \ref{sec:method}) and of the normalized standard error of the mean flux (usually of the order of $10$\% to $25$\%), evaluated based on the standard deviation and number of inferred fluxes. The corresponding error bars are provided (in red) in Figure \ref{fig1}. The difference between fluxes from the model and actual measured fluxes is usually less than a factor of $\sim1.5$, although it can sometimes increase to a factor of $\sim3$. One can also notice a rapid flux increase at low altitudes from 150 km to 2000 km, followed at $h>2000$ km by a slower increase well fitted by the model. 

Several metrics are used to assess the accuracy and the forecasting ability of the model. We calculate, for the training and the validation sets, MEF, $EF_{90}$ and the Pearson correlation coefficient $r$ between model values of $J_{omni}$ and values inferred from ELFIN measurements. We do this each time for the whole set and for each domain of $(AE_i^\star,L)$ and $(AE_i^{\star\star},L)$ separately. We also determine MEF and $EF_{90}$ at three altitudes, $h=150$ km, $2000$ km, and the maximum altitude $h\leq20,000$ km reached on the considered field line, for $E=100$ keV, $500$ keV, and 1.5 MeV, for each $AE_i^\star$ or $AE_i^{\star\star}$ level. 

\begin{table} 
\small
\begin{adjustwidth}{-2cm}{-2cm}
\centering
\setlength{\tabcolsep}{4pt}
\begin{tabular}{c c c c c c c c c c}
\makecell{ $AE$ \\ level} & \makecell{ $L$ \\ domain} &  $r$ &  MEF/$EF_{90}$ & \makecell{ MEF/$EF_{90}$ \\ (150 km)} & \makecell{ MEF/$EF_{90}$ \\ (2000 km)} & \makecell{ MEF/$EF_{90}$ \\ (max. alt.)} & \makecell{ MEF/$EF_{90}$ \\ (100 keV)} & \makecell{ MEF/$EF_{90}$ \\ (500 keV)} & \makecell{ MEF/$EF_{90}$ \\ (1.5 MeV)}\\
 \toprule
 {all}                 & 0 & 0.95 & 1.8/3.7 & 2.8/5.6 & 1.5/2.3 & 1.5/3.6 & 2.2/4.3 & 1.9/3.5 & 2.2/6.0 \\ \midrule \midrule
 {$AE_0^{\star}$}      & 1 & 0.93 & 1.6/2.5 & 1.9/3.0 & 1.5/2.0 & 1.3/1.8 & 1.4/1.6 & 1.7/3.2 & 1.7/3.0 \\  \midrule
 {$AE_1^{\star}$}      & 1 & 0.92 & 1.3/2.2 & 1.5/2.6 & 1.2/1.7 & 1.4/2.5 & 1.2/1.5 & 1.4/2.5 & 1.8/3.1 \\  \midrule
 {$AE_2^{\star}$}      & 1 & 0.87 & 1.4/2.3 & 1.7/3.3 & 1.3/2.0 & 1.4/2.2 & 1.4/1.9 & 1.4/2.3 & 1.9/5.1 \\  \midrule
 {$AE_3^{\star}$}      & 1 & 0.76 & 1.4/2.8 & 1.8/3.3 & 1.3/2.7 & 1.5/2.7 & 1.4/2.2 & 1.6/3.1 & 1.8/4.6 \\ \midrule \midrule
 {$AE_0^{\star\star}$} & 1 & 0.94 & 1.5/2.6 & 1.8/3.0 & 1.5/2.2 & 1.4/1.8 & 1.4/1.6 & 1.7/3.4 & 1.7/2.9 \\   \midrule
 {$AE_1^{\star\star}$} & 1 & 0.85 & 1.3/2.6 & 1.6/3.1 & 1.3/2.2 & 1.4/2.9 & 1.3/1.7 & 1.6/3.1 & 1.8/4.2 \\  \midrule
 {$AE_2^{\star\star}$} & 1 & 0.89 & 1.3/2.1 & 1.6/2.7 & 1.2/1.7 & 1.3/2.3 & 1.3/1.7 & 1.4/2.1 & 1.9/4.2 \\  \midrule
 {$AE_3^{\star\star}$} & 1 & 0.85 & 1.4/2.5 & 1.6/3.3 & 1.3/2.1 & 1.4/2.4 & 1.3/1.9 & 1.3/2.9 & 1.7/3.6 \\
\bottomrule
\end{tabular}
\caption{Accuracy metrics for the $J_{omni}(AE^\star)$ and $J_{omni}(AE^{\star\star})$ models}
\label{TablerrorsAEstarandAEdoublestar}
\end{adjustwidth}
\end{table}

Table \ref{TablerrorsAEstarandAEdoublestar} shows the performance of the $J_{omni}(AE^\star)$ and $J_{omni}(AE^{\star\star})$ models on the training dataset (ELFIN A). The global accuracy metrics for both the $AE^{\star}$ and $AE^{\star\star}$ models are nearly identical, with a Pearson correlation coefficient $r = 0.86$, a median error factor MEF $\sim1.4$, and a 90$^{th}$ percentile of the error factor $EF_{90}\sim2.5$ \blue{(equivalent to the $90$\% confidence interval)}. These metrics are usually roughly similar for the two models at all altitudes and electron energies, demonstrating the good accuracy of these models throughout the parameter domains, thanks to a large sample size within each domain.

A comparison of the model with the validation dataset (ELFIN B) shows a similarly good agreement, with Pearson correlation coefficients of $r=0.79$ and $r=0.83$, global MEFs of 1.5 and 1.4, and global Error Factors $EF_{90}$ of 3.1 and 3.0 for $AE^{\star}$ and $AE^{\star\star}$ models, respectively. Note that ELFIN A and B CubeSats often collect data from the same location (MLT sector) with a time lag of $\sim0.05-30$ minutes between the two spacecraft, which is sufficient to measure significantly different precipitating or trapped electron fluxes \cite{Zhang23:jgr:ELFIN&scales} and justifies using ELFIN B measurements as the validation set. 

\blue{As shown in many previous works, the measured electron fluxes in the Earth's outer radiation belt vary in time and space by factors of 10 to 10000 and, therefore, electron flux models (and even large numerical radiation belt models) cannot be expected to remain closer than a factor of $\approx2-3$ from measured fluxes at all times and locations \cite{Glauert14, Lima20, Sicard18, Smirnov20}. Therefore, median errors of a factor of $\sim1.5$, and maximum errors of a factor of $\sim2-3$ for 90\% of the data, between model fluxes and actual fluxes, as obtained here, can be considered as acceptable. The accuracy of the present flux model, obtained using a novel method, is similar to the accuracy of previous flux models obtained using different methods \cite{Glauert14, Lima20, Smirnov20}. }

\subsection{Omnidirectional Electron Flux Variations with MLT}
\label{subsec:MLT}

Previous investigations of electron fluxes measured by Polar Operational Environmental Satellites (POES) in polar orbit at 850 km altitude found a non-negligible MLT asymmetry of trapped electron flux up to at least $300$ keV at $L\sim3-9$ \cite{Allison17, Meredith16}, increasing with \blue{substorm activity} ($AE^\star$), with $\approx2$ times higher $100-300$ keV electron flux in the dawn sector than in the dusk sector at $L=3-7$ when $AE^\star<1000$ nT and the reverse at $L>7$ during quiet periods \cite{Allison17}. A dawn-dusk asymmetry of $200-350$ keV electron flux by a factor of $\approx2$ has also been observed at $L=1.3$ \cite{Selesnick16}.

Such MLT variations are partly due the MLT asymmetry of substorm electron injections from the plasmasheet and to non-dipolar components in the actual geomagnetic field. Electron injections usually occur at 0-6 MLT. During their subsequent azimuthal drift toward dusk, injected electrons are efficiently precipitated by whistler-mode waves (at $<500$ keV) or EMIC waves (at $>1$ MeV) into the atmosphere, or lost through magnetopause shadowing at $L>5$, leading to a lower electron flux in the dusk sector. A distortion of trapped electron drift shells by the solar wind-driven dawn-to-dusk convection electric field \cite{Matsui13} and ionospheric electric fields \cite{Lejosne16,Califf22} can also result in a dawn-dusk asymmetry in electron fluxes, a process which may operate down to $L=1.3$ \cite{Selesnick16}.

\begin{figure}
\centering
\includegraphics[width=\textwidth]{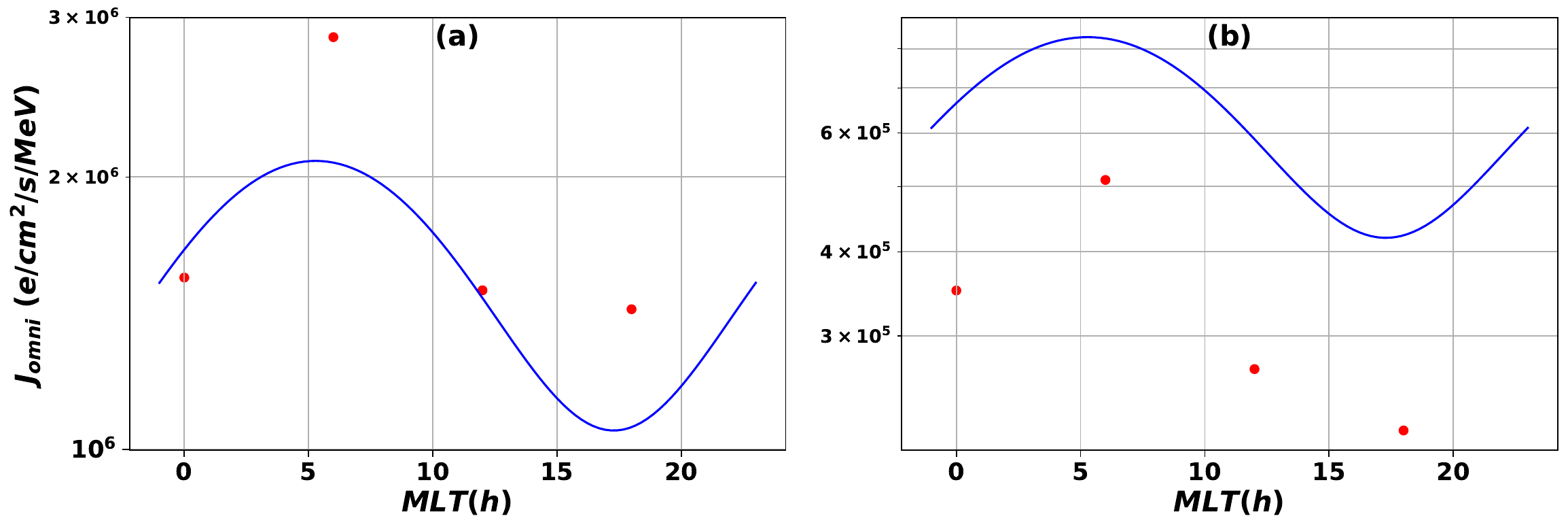}
\caption{(a) Model omnidirectional electron flux with included MLT modulation, $J_{omni}({\rm MLT}) = J_{omni} \times M({\rm MLT},K)$ at 100 keV, $L=5$, and $h=450$ km (in blue), as a function of MLT following periods of quiet geomagnetic conditions ($AE_1^\star$), with corresponding fluxes $J_{omni}({\rm MLT})$ directly inferred from ELFIN A measurements (in red). (b) Same as (a) for 300 keV electrons.}
\label{fig2}
\end{figure}

As in the above-discussed previous works, the time-averaged $J_{omni}(h,{\rm MLT})$ measured (at $h=h_0$) or inferred from ELFIN data in each MLT sector exhibits a non-negligible MLT modulation at all $E$ and $L$, usually by a factor of $\approx2$. To approximately take this MLT modulation into account, we minimize $J_{omni}({\rm MLT})/\langle J_{omni}\rangle - M({\rm MLT},K)$, where $M({\rm MLT},K) = 1 + 0.33 \sin(2\pi\,{\rm MLT}/24 + K)$ and the average is performed over MLT, giving us a $K$ value for each ($L$, $AE_i^\star$) or ($L$, $AE_i^{\star\star}$) domain, provided in the Appendix. Multiplying the MLT-averaged $J_{omni}(h)$ from the ELFIN-based model by the function $M({\rm MLT},K)$ allows to roughly incorporate MLT modulations. The resulting new version, with MLT modulation, of the ELFIN-based $J_{omni}$ model is however in slightly less good agreement with MLT-averaged $J_{omni}$ values inferred from ELFIN data than the baseline MLT-averaged model.
Figure \ref{fig2} shows 100 keV and 300 keV electron fluxes of the model with MLT modulation at $L=5$ and $h=450$ km (blue curve), compared with actual electron fluxes measured by ELFIN in different MLT sectors (red circles). \blue{Although there is a factor of $\sim1.4$ to $\sim2$ difference between the MLT-averaged flux level of the analytical model and the MLT-averaged flux level inferred from ELFIN in Figure \ref{fig2}, the relative increase by a factor of $\approx2$ of the inferred flux in the 6 MLT sector compared with the 0 MLT and $12-18$ MLT sectors is relatively well reproduced by the analytical model with MLT modulation. Note that we consider only four MLT sectors, to have a sufficient amount of data points in each MLT sector.}

\section{Analysis of Model Results}\label{sec:discussion}

\subsection{Electron Flux Variations with \blue{substorm activity}}

\begin{figure}
\centering
\includegraphics[width=\textwidth]{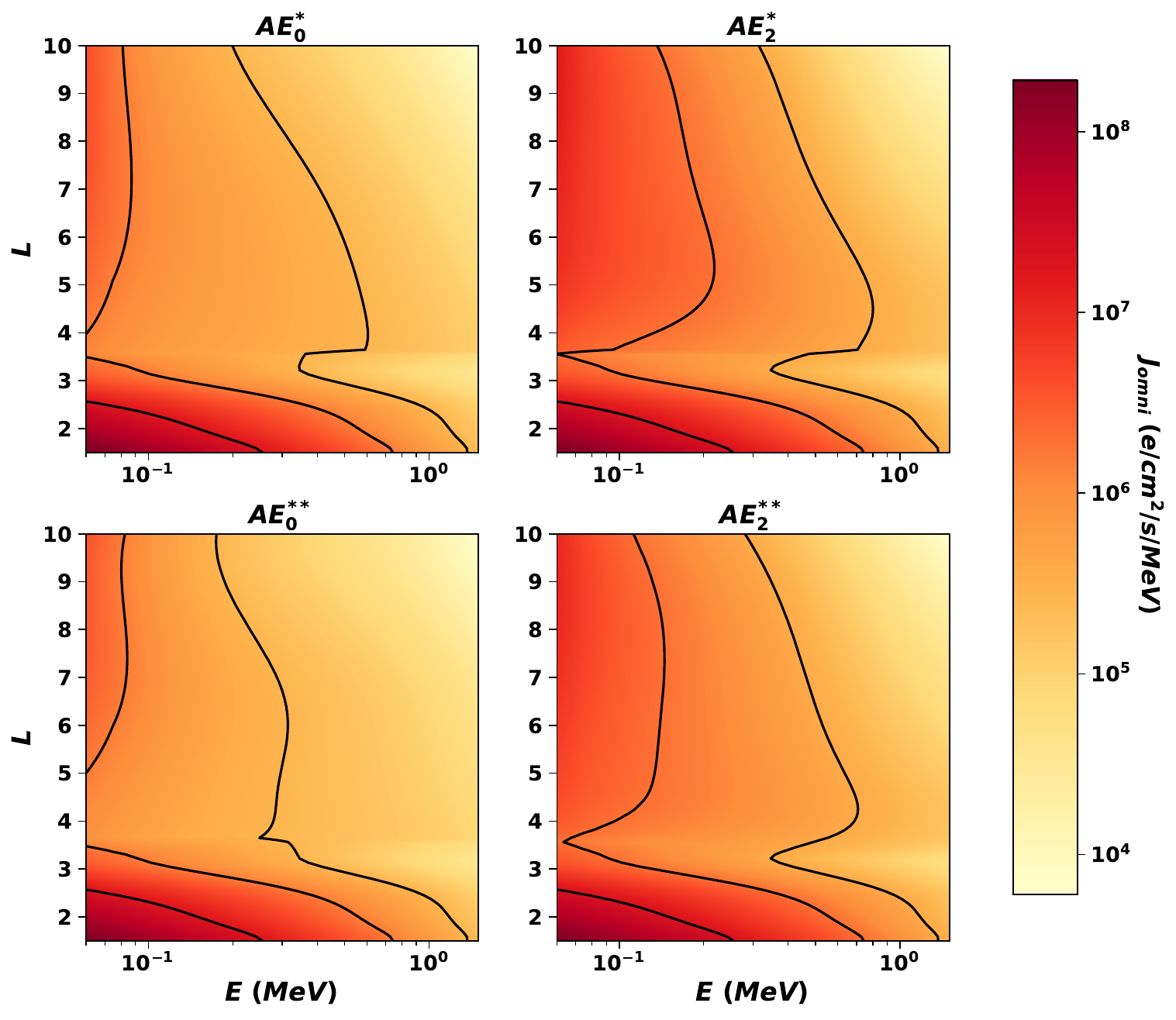}
\caption{Model average omnidirectional electron flux $J_{omni}(E,L,h)$ as a function of energy $E$ and $L$-shell at the altitude $h=h_0=450$ km of ELFIN A measurements, following periods of quiet (left) and disturbed (right) geomagnetic conditions, corresponding to $AE_0^\star$ and $AE_2^\star$ levels (top row) or $AE_0^{\star\star}$ and $AE_2^{\star\star}$ levels (bottom row). Black contours show $1/10$, $1/100$, and $1/500$ of the maximum flux in the color scale on the right-hand-side. }
\label{fig3}
\end{figure}

\begin{figure}
\centering
\includegraphics[width=\textwidth]{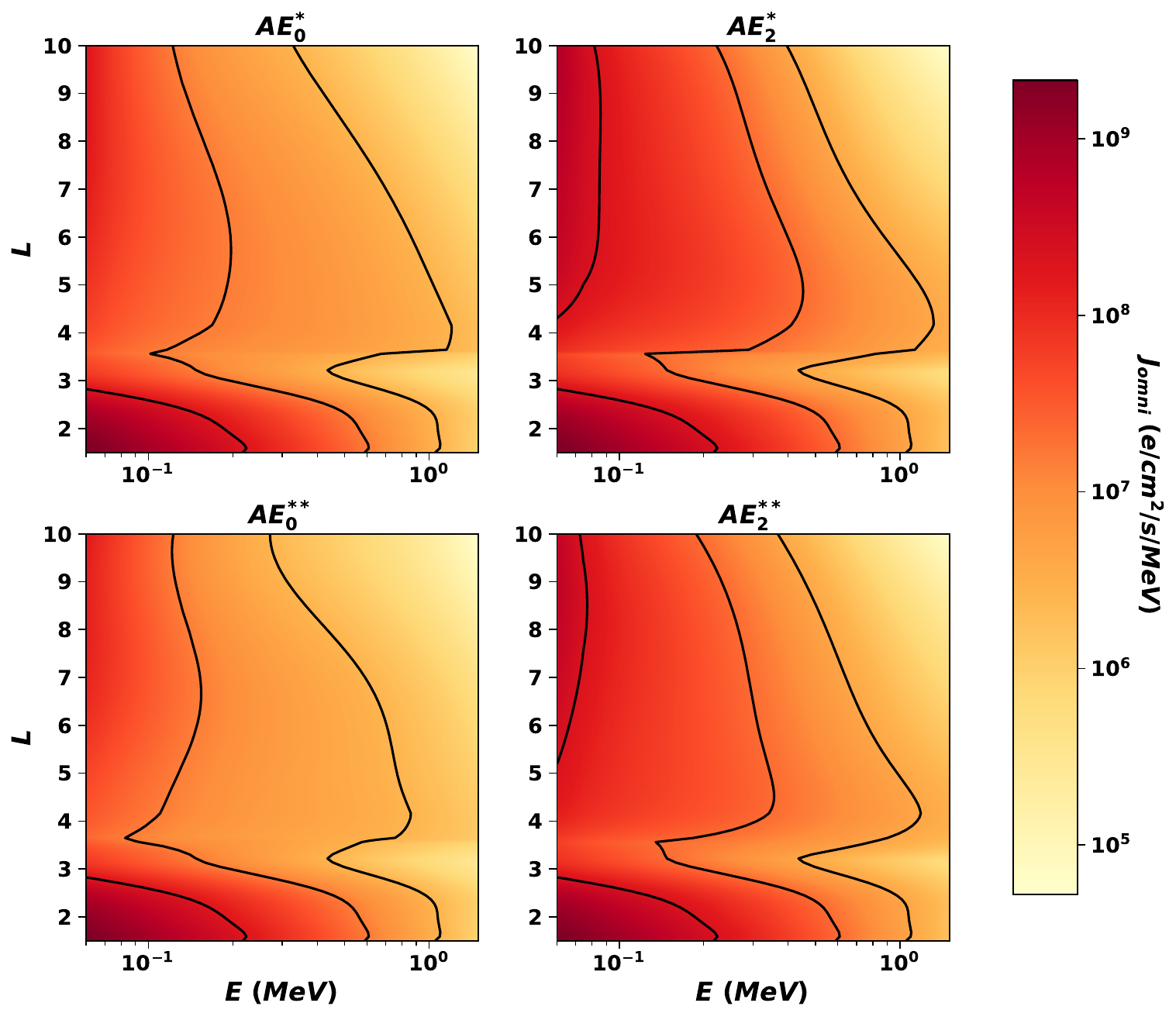}
\caption{Model average omnidirectional electron flux $J_{omni}(E,L,h)$ as a function of energy $E$ and $L$-shell at $h=\min(20,000$ km, $(L-1)\times 6371$ km), for quiet (left) and disturbed (right) geomagnetic conditions, corresponding to $AE_0^\star$ and $AE_2^\star$ levels (top row) or $AE_0^{\star\star}$ and $AE_2^{\star\star}$ levels (bottom row). Black contours show $1/10$, $1/100$, and $1/500$ of the maximum flux in the color scale on the right-hand-side.}
\label{fig4}
\end{figure}

Figures \ref{fig3} and \ref{fig4} show maps of the model omnidirectional electron flux $J_{omni}(E,L)$ for $L=1.5-10$ at low and high altitudes, $h=h_0=450$ km and $h=\min(20,000$ km, $(L-1)\times 6371$ km), respectively, following periods of quiet and disturbed geomagnetic conditions, defined by $AE_0^\star$ and $AE_2^\star$ levels (top row) or $AE_0^{\star\star}$ and $AE_2^{\star\star}$ levels (bottom row). The selected altitudes $h\leq h_{max}$ in Figure \ref{fig4} correspond to the magnetic equator from $L=1.5$ to $L=4.15$. Figures \ref{fig3} and \ref{fig4} demonstrate that the model describes well the structure of electron fluxes in the inner magnetosphere from $L=1.5$ up to $L=10$, with a first flux peak in the inner radiation belt at $L=1.5-2$, low fluxes in the slot region at $L\simeq3-3.5$ due to hiss wave-driven electron loss \cite{Lyons&Thorne73, Mourenas17}, high $0.3-1.5$ MeV electron fluxes in the outer radiation belt at $L\simeq4-7$, and the plasma sheet at $L>7$. Note that the present model provides only time-averaged fluxes at $L<3.25$, with exactly the same fluxes in left and right columns of Figures \ref{fig3} and \ref{fig4}. This is justified by the much weaker variation of time-averaged electron fluxes with geomagnetic activity at $L\lesssim3$ than at $L>3.5$ \cite{Mourenas17, Reeves16}.

The average omnidirectional flux is rapidly decreasing from 60 keV to 1.5 MeV at all $L$-shells, but less steeply within the outer radiation belt (at $L\simeq4-7$) than in the plasma sheet (at $L>7$). This is likely due to two physical processes: an adiabatic acceleration of electrons as they travel from $L\sim10$ to $L\sim4$ toward a stronger geomagnetic field \cite<partly through inward radial diffusion by ULF waves, see>[]{Ozeke14, Hudson21}, and an efficient local energization of $\sim100-500$ keV electrons by chorus waves in low plasma density regions of the outer radiation belt at $L=3.5-7$ \cite{Agapitov19:fpe, Horne05JGR, Summers98, Thorne13:nature}.

In the outer radiation belt, the radial ($L$) position of the maximum average omnidirectional $100-500$ keV electron flux comes closer to the Earth, down to $L\simeq4-4.5$, after periods of high \blue{substorm activity} (e.g., compare black contours of flux for $AE_0^{\star\star}$ and $AE_2^{\star\star}$ levels in Figure \ref{fig4}). The radial position of this maximum of $J_{omni}(L)$ after active periods is consistent with the position of the peak of chorus wave-driven electron energization inferred from statistical wave and plasma measurements, which similarly moves to lower $L$ after disturbed periods \cite{Agapitov19:fpe}. Equatorial measurements from the Van Allen Probes likewise show a maximum of omnidirectional $100-500$ keV electron flux at $L>5-6$ during quiet periods, moving to $L\simeq4-4.5$ during geomagnetic storms \cite{Reeves16}.

Following periods of high impulsive or time-integrated \blue{substorm activity} (corresponding to $AE_2^\star$ or $AE_2^{\star\star}$ levels, respectively), injections from the outer plasma sheet become more intense and both chorus and ULF wave power increase, leading to higher $J_{omni}$ at $L>3.5$ than after quiet periods (corresponding to $AE_0^\star$ or $AE_0^{\star\star}$ levels), at all energies. In the outer radiation belt, this increase of $J_{omni}$ is stronger at $L\simeq4-5$ in Figure \ref{fig4}. The electron flux above $200$ keV is initially low during quiet times at $L=4-10$, but it is lower for $AE_0^{\star\star}$ than for $AE_0^\star$, because electron flux measurements corresponding to $AE_0^\star$ (that is, with a mean $AE<100$ nT during the past 4 hours) are sometimes associated to significant time-integrated \blue{substorm activity} over the past 72 hours ($AE_1^{\star\star}$ or $AE_2^{\star\star}$ levels) that have already increased electron fluxes. Indeed, electron energization at $L\simeq4-5$ is often a cumulative process, requiring many hours of elevated \blue{substorm activity} \cite{Hua22b, Mourenas19:Impact, Mourenas22:jgr:climatology, Thorne13:nature}. Figure \ref{fig4} shows that whatever the \blue{substorm activity} during the preceding 4 hours ($AE^\star$), the level of $J_{omni}$ will also partly depend on the level of time-integrated \blue{substorm activity} over the preceding 3 days ($AE^{\star\star}$).

Therefore, Figure \ref{fig4} demonstrates the important role of time-integrated \blue{substorm activity} ($AE^{\star\star}$) in controlling the average omnidirectional electron flux over a very wide parameter range, from $\sim60$ keV to $1.5$ MeV and from $L=3.5$ to $L=10$. This suggests that the build-up of energetic electron fluxes often takes place over many consecutive hours in the near-Earth plasma sheet, even before reaching the outer radiation belt, probably through progressive convection and betatron acceleration within dipolarizing flux bundles at $L\sim9-15$ \cite<e.g., see>[and references therein]{Gabrielse17}, as well as through electron inward radial diffusion and chorus wave-driven energization closer to the Earth \cite{Mourenas22:jgr:climatology, Ozeke14, Simms21, Thorne13:nature}. Nevertheless, the significant increase of electron fluxes with $AE^\star$ in Figure \ref{fig4} indicates a strong concomitant influence of impulsive events in shaping energetic and relativistic electron fluxes at $L=3.5-10$.

\subsection{Comparisons with Van Allen Probes, THEMIS, and POES Data}\label{sec:comparisons}

\begin{figure}
\centering
\includegraphics[width=\textwidth]{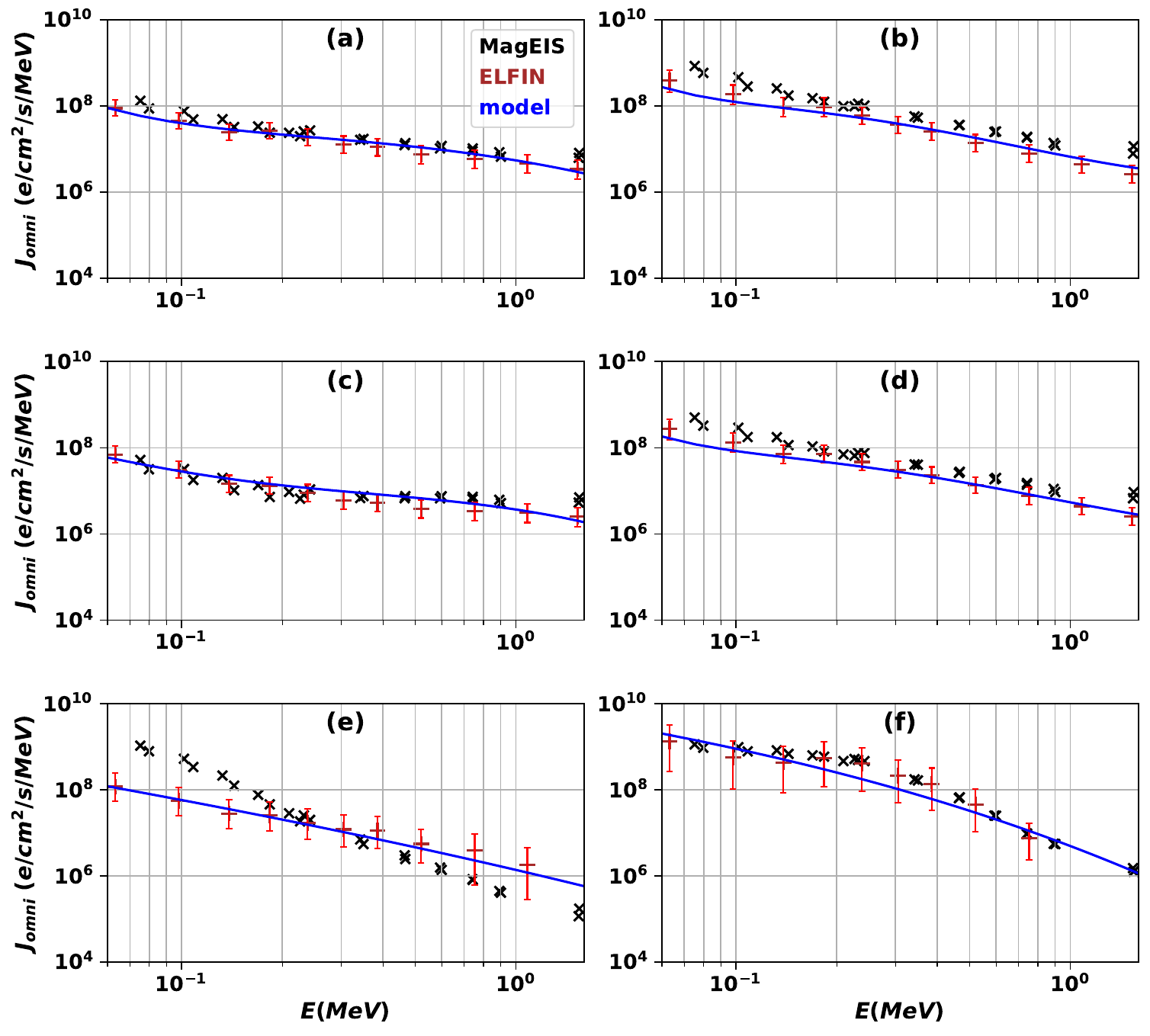}
\caption{(a) Model average $J_{omni}(E)$ at $L\approx4.5$ and $h=22,000$ km for the low \blue{substorm activity} level $AE_0^\star$ (blue curve), corresponding average flux inferred from ELFIN data (red pluses with error bars), and corresponding average omnidirectional electron flux from MagEIS onboard the Van Allen Probes near the magnetic equator in 2017-2018 (black crosses). (b,c,d) Same as (a) for \blue{substorm activity} levels $AE_2^\star$, $AE_0^{\star\star}$, and $AE_2^{\star\star}$, respectively. (e) Same as (a) at $L\approx3.0$ and $h=12,600$ km but averaged over all $AE$ domains. (f) Same as (e) at $L\approx1.5$ and $h=3,150$ km.}
\label{fig5}
\end{figure}

The average $J_{omni}$ from the present model (blue line) is compared in Figure \ref{fig5} to the corresponding average flux inferred from ELFIN data (red pluses with error bars) and to the average omnidirectional electron flux measured by the Magnetic Electron Ion Spectrometer (MagEIS) aboard the Van Allen Probes \cite{Blake13, Claudepierre21} in 2017-2018 less than $10^\circ$ from the magnetic equator (black crosses). We use MagEIS data from 2017-2018, because this period, like the 2020-2022 period of ELFIN measurements, took place within 3 years of the solar cycle minimum of December 2019, suggesting roughly similar space weather properties during these two periods. \blue{We checked that substorm activity was indeed very similar in 2017-2018 and 2020-2022, with very similar probability distribution functions (PDF) of $AE^\star$ and $AE^{\star\star}$, similar average $\langle AE^\star\rangle$ ($\simeq440$ nT and $470$ nT, respectively) and standard deviation $\sigma(AE^\star)$ ($\simeq340$ nT and $380$ nT, respectively), and similar average $\langle AE^{\star\star}\rangle$ ($\simeq1.5\times10^4$ nT$\cdot$h and $1.7\times10^4$ nT$\cdot$h, respectively) and standard deviation $\sigma(AE^{\star\star})$ ($\simeq8\times10^3$ nT$\cdot$h in both cases). Nevertheless, there were also twice more frequent periods of $Kp\geq5$ and $Kp\geq6$ in 2017-2018 than in 2020-2022, and three large geomagnetic storms (with $\min(Dst)=-146$ nT to $-176$ nT) in 2017-2018 versus one large storm (with $\min(Dst)=-105$ nT) in 2020-2022, which could have led to higher time-averaged $\sim60-200$ keV electron fluxes at $L=2.5-3.5$ in 2017-2018 than in 2020-2022 \cite{Califf22, Mei23, Mourenas17, Turner17b, Zhao23}.}

Electron fluxes are displayed at $L\approx4.5$ and $h=22,000$ km in Figures \ref{fig5}(a)-\ref{fig5}(d) for \blue{substorm activity} levels $AE_0^\star$, $AE_2^\star$, $AE_0^{\star\star}$, and $AE_2^{\star\star}$. Time-averaged fluxes (averaged over all $AE$ levels) from the model at $L\approx3.0$ and $h=12,600$ km and at $L\approx1.5$ and $h=3,150$ km are also displayed in Figures \ref{fig5}(e) and \ref{fig5}(f), respectively, with corresponding average fluxes inferred from ELFIN data and measured by the Van Allen Probes in 2017-2018. Thanks to numerous data, the typical standard normalized error on average fluxes from the Van Allen Probes is only 4.5\% (it is always less than 8.5\%).

The adopted altitudes $h(L)$ for model and inferred fluxes in Figure \ref{fig5} correspond to the magnetic equator, as required for comparisons with the bulk of the Van Allen Probes measurements. As average fluxes are inferred from ELFIN data at the same 18 pre-determined altitudes $h_n$ for all $L$, a slight logarithmic extrapolation is performed to obtain values at $h(L)$ displayed in Figure \ref{fig5}, using $\ln(J_{omni}(h(L))/J_{omni}(h_n))/\ln(J_{omni}(h_{n-1})/J_{omni}(h_n)) = \ln(h(L)/h_n)/\ln(h_{n-1}/h_n)$, where $h_n$ and $h_{n-1}$ are the two closest altitudes below $h(L)$ where inferred fluxes are available. Note, however, that $h(L)$ is slightly higher than $h_{max}(L)$ given by Equation (\ref{eq6}) at $L=4.5$, which may lead to larger discrepancies between model fluxes and actual fluxes than at $h<h_{max}$ (see Section \ref{sec:method}). In addition, the present $J_{omni}$ model has been fitted to fluxes inferred from ELFIN data only at $h\leq h_n\leq20,000$ km. But the slow increase with $h$ of both modelled and inferred $J_{omni}(h)$ at $h>4,000$ km (e.g., see Figure \ref{fig1}) suggests that the model should remain approximately valid up to $h=22,000$ km at $L=4.5$.

\begin{figure}
\centering
\includegraphics[width=\textwidth]{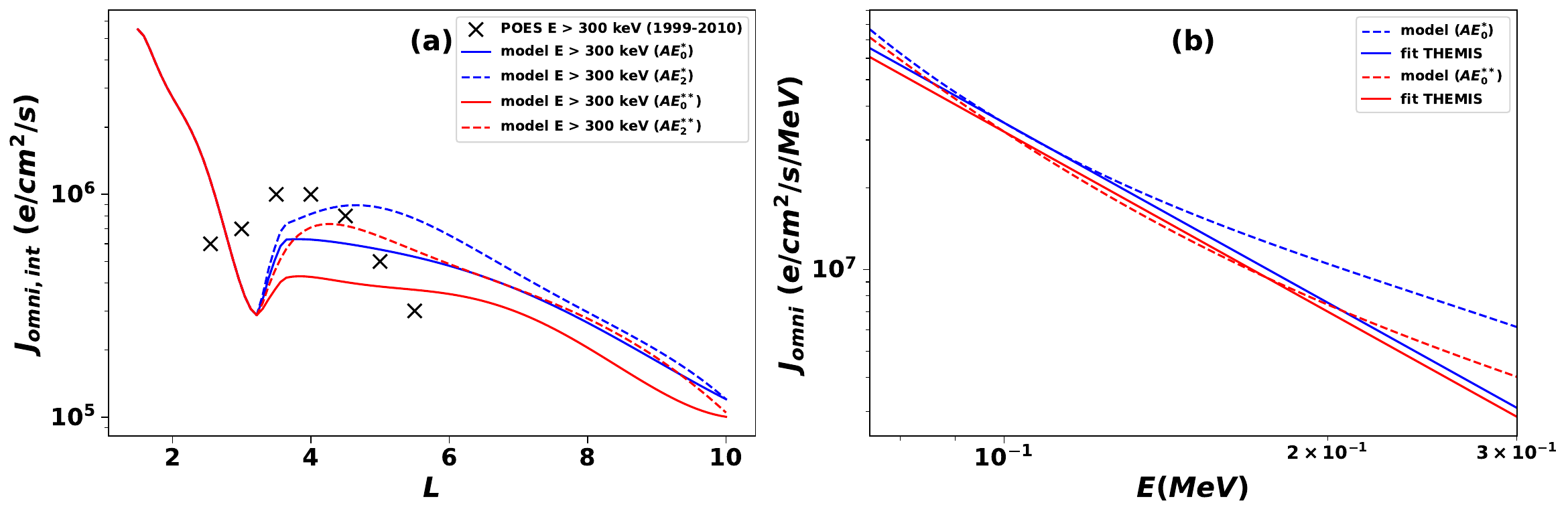}
\caption{(a) Average integral omnidirectional electron flux from the present model at $E>300$ keV for $h\simeq850$ km and \blue{substorm activity} levels $AE_0^\star$ and $AE_2^\star$ (blue solid and dashed curves, respectively), as well as $AE_0^{\star\star}$ and $AE_2^{\star\star}$ (red solid and dashed curves, respectively) as a function $L$, with corresponding average fluxes from POES in 1999-2010 (black crosses). (b) Model $J_{omni}(E)$ at $L=9$ and $h=20,000$ km for \blue{substorm activity} levels $AE_0^\star$ and $AE_0^{\star\star}$ (dashed blue and red curves, respectively), and normalized power-law fits $J_{omni}(E)/J_{omni}(100\,{\rm keV})\approx(100\,{\rm keV}/E)^\alpha$ with $\alpha=2.2$ (solid blue and red lines) corresponding to typical omnidirectional $75-300$ keV electron flux energy spectra measured by THEMIS spacecraft at $L\approx9$ around the magnetic equator in 2008-2020. }
\label{fig6}
\end{figure}

Figure \ref{fig5} shows that the average electron flux of the model usually remains close to the average flux measured in 2017-2018 by the Van Allen Probes near the magnetic equator, from $L=1.5$ to $L=4.5$ over the whole parameter range. Global metrics for the $AE^\star$ and $AE^{\star\star}$ models compared to Van Allen Probes fluxes are median error factors MEF $=2.1$ for both \blue{substorm activity} indicators, with $90^{th}$ percentiles of the error factor $EF_{90}=6.0$ and $EF_{90}=6.9$, and Pearson correlation coefficient $r=0.60$ for both. Notably, the increase of electron fluxes from the Van Allen Probes with \blue{substorm activity} at $100-900$ keV is well reproduced by the model at $L=4.5$. The average electron flux of the model decreases faster at $L=1.5-3$ than at $L=4.5$ from $100$ keV to $1.5$ MeV, and it is much lower in the slot region at $L=3$ than in both the inner radiation belt at $L=1.5$ and the outer radiation belt at $L=4.5$ above $100$ keV, in agreement with observations from the Van Allen Probes. The slot region is produced by hiss wave-driven electron precipitation into the atmosphere \cite{Lyons&Thorne73, Ma16:hiss, Mourenas17}. 

The comparisons in Figure \ref{fig5} therefore provide a validation of the present method for inferring omnidirectional electron fluxes at all altitudes along a geomagnetic field line based on low-altitude ELFIN data of trapped, precipitating, and backscattered electron fluxes. The largest discrepancy between model and Van Allen Probes fluxes occurs at $70-150$ keV and $L=3$, where average fluxes from the model are $\approx4-8$ times lower than Van Allen Probes fluxes. This could partly reflect actual differences between electron fluxes in 2020-2022 and 2017-2018. \blue{There were indeed twice more frequent periods of $Kp\geq5$ and $Kp\geq6$ in 2017-2018 than in 2020-2022, which could have led to a higher time-averaged flux of energetic $\sim60-200$ keV electrons at $L\simeq3$ in 2017-2018 than in 2020-2022, due to deep electron penetrations down to $L\simeq3$, which are much more frequent at $E<250$ keV \cite{Turner17b, Zhao23} and mainly occur during periods of high $Kp>5-6$ due to enhanced convection or Subauroral Polarization Streams (SAPS) electric fields \cite{Califf22, Mei23, Zhao23}.} But this discrepancy is probably also due to peculiarities of wave-driven electron pitch-angle diffusion rates at $L\approx3$, which sometimes exhibit a bottleneck (a deep trough) at moderately high pitch-angles and low energy, leading to much larger fluxes than in the present model above $\alpha_{eq}=60^\circ$ (see Section \ref{sec:method}). 

Figure \ref{fig6}(a) further shows the average energy-integrated omnidirectional electron flux at $E>300$ keV from the present ELFIN-based model at $h\simeq850$ km, for \blue{substorm activity} levels $AE_0^\star$ and $AE_2^\star$ (blue solid and dashed curves, respectively) as well as $AE_0^{\star\star}$ and $AE_2^{\star\star}$ (red solid and dashed curves, respectively), and the average $>300$ keV electron fluxes measured by Polar Operational Environmental Satellites (POES) in polar orbit at $\sim850$ km altitude \cite{Yando11}. Only average POES fluxes measured in 1999-2010 (black crosses) at adiabatically invariant shells $L^\star\leq5.5$ \cite{Sicard18} are displayed, because deviations of $L^\star$ from $L$ usually remain moderate in this domain, allowing us to assume that $L\approx L^\star$ to first order. \blue{Figure \ref{fig6}(a) shows that the time-averaged fluxes from the present model remain in reasonable agreement at $L\approx2.5-5.5$ with time-averaged fluxes from POES, despite the fact that such POES measurements were performed during another, more distant solar cycle, roughly $\sim10-20$ years before ELFIN measurements.}

Finally, Figure \ref{fig6}(b) shows that the model energy spectra $J_{omni}(E)$ at $L=9$ and $h=20,000$ km for low \blue{substorm activity} levels $AE_0^\star$ (dashed blue curve) and $AE_0^{\star\star}$ (dashed red curve) are in good agreement with the typical (median) energy spectrum shape $J_{omni}(E)/J_{omni}(100\,{\rm keV})\approx(100\,{\rm keV}/E)^\alpha$ with $\alpha=2.2\pm0.5$ (solid blue and red lines) of omnidirectional $75-300$ keV electron fluxes measured by Time History of Events and Macroscale Interactions during Substorms (THEMIS) spacecraft \cite{Angelopoulos08:sst} at $L\approx9\pm2$ around the equator (corresponding to fluxes higher than model fluxes) in the near-Earth plasma sheet in 2008-2020 \cite{Gao23}. At $L\approx9$, an altitude $h=20,000$ km is far from the magnetic equator, but while the absolute level of $J_{omni}$ should decrease as latitude increases along closed magnetic field lines, one expects a roughly similar variation of $J_{omni}(E)$ with energy $E\in[75,300]$ keV at different altitudes $h\gtrsim 20,000$ km when these electrons are sufficiently strongly scattered in pitch-angle to reach $h=450$ km in large numbers \cite{Gao22:Luphi_whistlers, Mourenas24:jgr:ELFIN&KPlimit, Shane23}. 

\subsection{Corresponding Sinusoidal Equatorial Pitch-Angle Distributions}

\begin{figure}
\centering
\includegraphics[width=\textwidth]{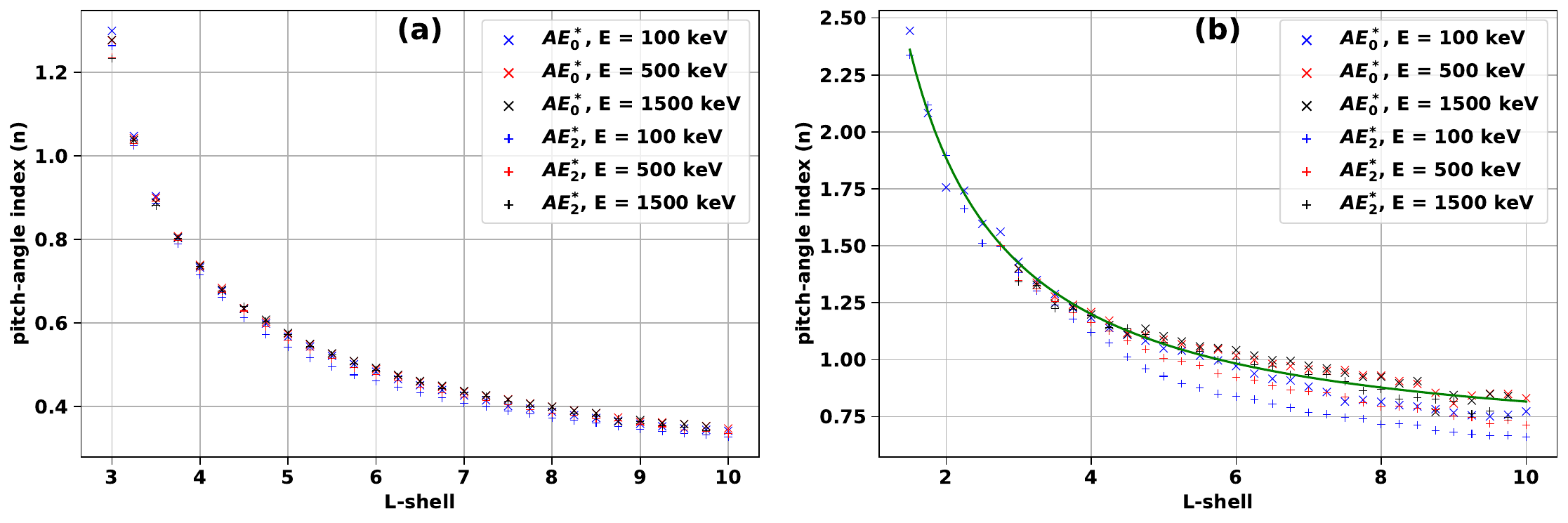}
\caption{(a) Pitch-angle index $n$ estimated between $\alpha_{eq}=10^\circ$ and $\alpha_{eq}=90^\circ$ based on the average precipitating to trapped flux ratio $J_{prec}/J_{trap}$ measured by ELFIN, as a function of $L$ for 100 keV (blue), 500 keV (red), and 1.5 MeV (black) electrons, following quiet ($AE_0^\star$) and disturbed ($AE_2^\star$) periods (crosses and pluses, respectively). (b) Same as (a) but showing $n$ estimated between $\alpha_{eq}=1.1\times \alpha_{eq,LC}$ and $\alpha_{eq}=90^\circ$, with a best fit $n=0.59+0.22\times (10/L)^{1.09}$ to $AE_0^\star$ model values (solid green curve).}
\label{fig7}
\end{figure}

The pitch-angle distribution of electrons in the Earth's radiation belts has been examined in many past studies, often using simple fits of the form $J(\alpha_{eq}) = \sin^n\alpha_{eq}$, with $n$ the pitch-angle index \cite{Gannon07, Shi16, Zhao14}. Therefore, it is interesting to compare the new ELFIN-based model, based on low-altitude measurements with good resolution at low equatorial pitch-angles, with these past results derived from spacecraft measurements at high altitude around the magnetic equator with a low resolution at low equatorial pitch-angles but a good resolution at high equatorial pitch-angles. Figure \ref{fig7} shows the pitch-angle index $n$ calculated using Equations (\ref{eq4}) and (\ref{eq5}) under the assumption that $J(\alpha_{eq})=\sin^n\alpha_{eq}$ based on the average precipitating to trapped flux ratio $J_{prec}/J_{trap}$ measured at low altitude by ELFIN, for $0.1$, $0.5$, and 1.5 MeV at $L=1.5-10$, for quiet ($AE_0^\star$) and active ($AE_2^\star$) geomagnetic conditions. Figure \ref{fig7}(a) shows a first estimate of $n$ calculated between $\alpha_{eq}=10^\circ$ and $\alpha_{eq}=90^\circ$. Figure \ref{fig7}(b) shows a second estimate of $n$ calculated between $\alpha_{eq}=1.1\times \alpha_{eq,LC}$ and $\alpha_{eq}=90^\circ$. These two estimates of $n$ are expected to be roughly similar to estimates obtained in previous works by fitting the measured equatorial pitch-angle electron distribution at $\alpha_{eq}> 5^\circ-15^\circ$. The second $n$ estimate may also allow inferring $J(\alpha_{eq}=90^\circ)$ from $J(\alpha_{h_0}\simeq90^\circ)$ measured by ELFIN. A best fit $n=0.59+0.22\times (10/L)^{1.09}$ (solid green curve) to this second estimate for $AE_0^\star$ periods is shown in Figure \ref{fig7}(b).

For $0.1$ MeV, $0.5$ MeV, and $1.5$ MeV electrons, Figure \ref{fig7} shows that the pitch-angle index $n$ inferred from ELFIN low-altitude measurements decreases as $L$ increases, from $n\simeq1.75-2.5$ at $L=1.5-2.5$ to $n\simeq0.35-1.25$ at $L=4-10$, with a slightly lower $n$ at lower energy or after disturbed periods ($AE_2^\star$). This decrease of $n$ toward higher $L$, lower $E$, or after stronger \blue{substorm activity}, is due to a stronger wave-driven pitch-angle diffusion that results in a flatter pitch-angle distribution and a higher $J_{prec}/J_{trap}$ \cite{Agapitov18:jgr, Kennel&Petschek66, Mourenas24:jgr:ELFIN&KPlimit}. These results are in good agreement with previous works based on near-equatorial electron fluxes measured by the Van Allen Probes, which obtained $n\approx2-4$ at $L=1.5-2.5$ and $n\simeq0.5-1.2$ at $L=3.5-6$ for $0.1-1$ MeV electrons and a slightly lower $n$ at lower energy \cite{Allison18, Shi16, Zhao14}, as well as a slightly lower $n$ after disturbed periods at $L=4-5$ \cite{Olifer22}. However, these previous works also found high indices $n\geq5$ for $100-350$ keV at $L\simeq3$, contrary to the present results. The good agreement between average fluxes $J_{omni}$ from the ELFIN-based model and average fluxes from Van Allen Probes 2017-2018 measurements near the magnetic equator at $L=3$ for $200-350$ keV in Figure \ref{fig5}(e) suggests that this discrepancy is probably due to a dominant contribution, in time-averaged fluxes, from high fluxes characterized by $n\approx1.5$ at $L=3$, while higher indices $n>5$ correspond to more frequent but much lower fluxes \cite{Shi16}. 

\section{Conclusions}\label{sec:conclusions}

In the present work, energy- and pitch-angle-resolved precipitating, trapped, and backscattered electron fluxes measured at 450 km altitude by ELFIN CubeSats in 2020-2022 have been used to infer omnidirectional fluxes of 60 keV to 1.5 MeV electrons at $L=1.5-10$, from 150 km up to $20,000$ km altitude, using adiabatic transport theory and quasi-linear diffusion theory. The inferred fluxes have been fitted by analytical formulas, using stepwise multivariate optimization. The resulting self-consistent model of omnidirectional electron flux $J_{omni}$, inferred from measurements from only one spacecraft at a time in LEO, is free from potential conjunction or intercalibration problems arising when combining measurements from different spacecraft. The present modelled electron fluxes are intrinsically coherent at all altitudes along each geomagnetic field line. Such modelled electron fluxes are provided as a function of electron energy $E$, $L$-shell, altitude, and of two different indices, $AE^\star$ and $AE^{\star\star}$, of \blue{substorm activity} over the preceding 4 hours and over the previous 72 hours, respectively. 

\blue{The present analytical flux model is valid on closed magnetic field lines, where electrons remain stably trapped. In practice, this implies that the model should be valid at all MLTs up to $L\simeq9-10$ when $Kp\leq4$ or $AE\leq700$ nT, and up to at least $L\simeq6-7$ during strong storms and substorms with $Kp>4$ and $AE>700$ nT at times when $Dst>-100$ nT \cite{Olifer18, Roederer18}. In addition, the present model should be valid near the magnetic equator only up to $L\approx4.5$, whereas at $L>4.5$ it should remain approximately valid only sufficiently far from the magnetic equator, corresponding to altitudes $h\lesssim 20,000$ km.}

The obtained model of omnidirectional electron flux has been validated by comparisons with Van Allen Probes measurements near the magnetic equator at high altitudes and $L\simeq1.5-4.5$ in 2017-2018, THEMIS measurements at $L\approx9$ in 2008-2020, and POES measurements at $850$ km altitude and $L\simeq2.5-5.5$ in 1999-2010. A version of the model includes MLT modulations similar to previous observations. The equatorial pitch-angle electron distributions corresponding to the present ELFIN-based model are also in rough agreement with previous works. Such comparisons show that the present method for inferring omnidirectional electron fluxes at all altitudes along a given magnetic field line based only on pitch-angle and energy resolved electron flux measurements from one low-altitude spacecraft is essentially valid from $L\sim1.5$ to $10$ above 60 keV, at least up to an altitude of $\approx20,000$ km. This study therefore suggests that a fleet of low-cost CubeSats similar to ELFIN could be used to monitor the radiation belts from low Earth orbit via the present method \cite{Millan24}.

The obtained model of omnidirectional electron flux allowed us to show that both impulsive \blue{substorm activity} ($AE^\star$) and time-integrated \blue{substorm activity} ($AE^{\star\star}$) are partly controlling the average level of $60-1500$ keV electron fluxes over a very wide $L$-shell domain, extending from $L=3.5$ to $L=10$. This new analytical model can be used for various tasks: (a) to obtain event-specific boundary conditions for radiation belt numerical models \cite<e.g., see>[]{Tu14}, (b) to assess the main physical processes at work through an examination of the latitudinal distribution of electron fluxes along field lines, (c) to disentangle the effects of rapid processes (such as direct electron injections) from the effects of slower cumulative processes (such as electron inward radial diffusion and wave-driven diffusive energization) by taking advantage of the functional dependence of modelled fluxes on two different \blue{substorm activity} indices (corresponding to brief and prolonged processes, respectively), and (d) to estimate or forecast the radiation dose along a given orbit and the related internal or surface charging hazards for satellites. Since the modelled $J_{omni}$ is provided in two separate versions, $J_{omni}(AE^\star)$ and $J_{omni}(AE^{\star\star})$, better estimates or forecasts of the omnidirectional electron flux may also be obtained by using an average of $J_{omni}(AE^\star)$ and $J_{omni}(AE^{\star\star})$, based on three-day sequences of measured or predicted $AE$ values. \blue{The proposed flux model could probably be improved by additionally taking $Dst$ or $Pdyn$ into account, but this would make it more complex, and this is left for future work.}

\acknowledgments
A.V.A., X.J.Z., and V.A. acknowledge support by NASA awards 80NSSC20K1270, 80NSSC23K0403, 80NSSC24K0558, NAS5-02099, and NSF grants AGS-1242918, AGS-2019950, and AGS-2021749. We are grateful to NASA's CubeSat Launch Initiative for ELFIN's successful launch. We acknowledge early support of the ELFIN project by the AFOSR, under its University Nanosat Program; by the UNP-8 project, contract FA9453-12-D-0285; and by the California Space Grant program. We sincerely acknowledge the critical contributions of the numerous volunteer ELFIN team student members.  

\subsection*{Open Research}
\noindent Electron fluxes measured by ELFIN are available in CDF format \cite{elfin_data}. Van Allen Probes MagEIS electron flux data (REL03 L2) is available from the New Mexico Consortium \cite{RBSP_data}. The $\mathit{SME}$ index \cite{Gjerloev12} is available at the SuperMAG data archive \cite{SuperMag}. OMNI data of $Dst$ and $Kp$ are available from the Kyoto World Data Center for Geomagnetism \cite{Geomagnetic_Indices_WDC}. Data access and processing was done using SPEDAS V3.1, see \citeA{Angelopoulos19}. 

\newpage
\appendix
\section{Model coefficients without/with MLT modulation}

The coefficients $B$ and $C_m$ in Equations (\ref{eq7}) and (\ref{eq8}) for the MLT-averaged model are provided in each parameter domain in Table \ref{TablecoeffsAEstarBandC} for both $AE^{\star}$ and $AE^{\star\star}$ models. The additional coefficients $K$, allowing to incorporate a MLT modulation in the MLT-averaged model (see Section \ref{subsec:MLT}), are provided in Table \ref{tableK} in each parameter domain.

\begin{table}[H]
\small
\begin{adjustwidth}{-3cm}{-3cm}
\setlength{\tabcolsep}{3pt}
\centering
\begin{tabular}{c c c c c c c c c c c c c}
 \makecell{$AE$ \\ level} &  \makecell{$L$\\ domain} &\bfseries $B$ &\bfseries $C_0$ & \bfseries $C_1$ & \bfseries $C_2$ & \bfseries $C_3$ & \bfseries $C_4$ & \bfseries $C_5$ & \bfseries $C_6$ & \bfseries $C_7$ & \bfseries $C_8$ & \bfseries $C_9$\\
 \hline
 all              & 0 & 0.705 & -590.88 & 282.36 & -65.015 & 5.830 & 634.4 & -261.5 & -0.603 & N/A & 0.0251 & -0.00523 \\ \hline
 {$AE_0^{\star}$} & 1 & 0.372 & -17.68 & -11.05 & -0.382 & 0.0119 & 70.1 & -77.6 & 6.413 & 18.9 & -0.0502 & -0.254 \\ \hline
 {$AE_1^{\star}$} & 1 & 0.391 & 3.79 & -88.83 & -0.145 & 0.00519 & 4.34 & -3.72 & 64.160 & 61.9 & 0.238 & -14.34 \\ \hline
 {$AE_2^{\star}$} & 1 & 0.462 & 84.94 & -209.59 & 1.009 & -0.0314 & -218.7 & 231.5 & 147.833 & 122.0 & 0.670 & -34.96 \\ \hline
 {$AE_3^{\star}$} & 1 & 0.436 & -6.86 & -153.65 & 0.0110 & -0.00303 & 75.6 & -130.1 & 111.192 & 103.5 & 0.424 & -25.04 \\ \hline
 \hline
 {$AE_0^{\star\star}$} & 1 & 0.304 & -87.24 & 111.16 & -1.390 & 0.0443 & 262.4 & -283.5 & -75.879 & -49.3 & -0.421 & 19.10 \\ \hline
 {$AE_1^{\star\star}$} & 1 & 0.413 & 33.36 & -196.69 & 0.417 & -0.0143 & -59.1 & 40.6 & 141.199 & 123.6 & 0.599 & -32.70 \\ \hline
 {$AE_2^{\star\star}$} & 1 & 0.451 & -35.33 & -136.60 & -0.326 & 0.00614 & 158.2 & -221.3 & 105.545 & 90.3 & 0.458 & -24.66 \\ \hline
 {$AE_3^{\star\star}$} & 1 & 0.528 & 113.16 & -149.56 & 0.935 & -0.0225 & -348.4 & 428.2 & 96.831 & 90.7 & 0.368 & -21.78 \\ \hline
\end{tabular}
\caption{$B$ and $C$ coefficients associated to $AE^\star$ or $AE^{\star\star}$}
\label{TablecoeffsAEstarBandC}
\end{adjustwidth}
\end{table}

\begin{table}[H]
\small
\centering
\begin{tabular}{c c c c}
\makecell{ $AE$ level} & \makecell{ $L$ domain} & $K_{AE^\star}$ &$K_{AE^{\star\star}}$\\
 \toprule
 all                                  & 0 & 2.68 & 2.68 \\ \midrule
 {$AE_0^\star$ or $AE_0^{\star\star}$} & 1 & 1.83 & 1.64 \\ \midrule
 {$AE_1^\star$ or $AE_1^{\star\star}$} & 1 & 0.19 & 5.80 \\ \midrule
 {$AE_2^\star$ or $AE_2^{\star\star}$} & 1 & 0.16 & 6.11 \\ \midrule
{$AE_3^\star$ or $AE_3^{\star\star}$} & 1 & 5.52 & 0.57 \\
\bottomrule
\end{tabular}
\captionsetup{justification=centering}
\caption{Coefficients $K$ for the MLT correction to models associated to $AE^\star$ and $AE^{\star\star}$}
\label{tableK}
\end{table}

%% ---------------------------------------------------------------%%
%% References and Citations %%
\newpage
%\input{appendices}

%\bibliography{full,addon}

\end{document}